\PassOptionsToPackage{table}{xcolor}
\documentclass[10pt,sigconf,letterpaper,nonacm]{acmart}
\usepackage{xcolor}

\usepackage{tabularray}

\usepackage{soul}
\usepackage{multirow}
\usepackage{caption}
\usepackage{subcaption}
\usepackage[colorinlistoftodos]{todonotes}

\usepackage{minibox}
\usepackage{algorithm}
\usepackage{algpseudocode}

\usepackage{appendix}
\newcounter{observcntr}
\newcommand*{\observ}[1]{%
    \stepcounter{observcntr}%
    \begin{center}
    \vspace{2pt}
    \minibox[frame, rule=1pt,pad=3pt]{
        \begin{minipage}[t]{0.95\columnwidth}
        \textbf{Answer RQ~\arabic{observcntr}:} \textit{#1}.
        \end{minipage}
    }
    \vspace{2pt}
    \end{center}
}

\definecolor{pinky}{RGB}{1,0.8,0.788}

\definecolor{orangy}{RGB}{1,0.8,0.404}

\AtBeginDocument{%
  \providecommand\BibTeX{{%
    \normalfont B\kern-0.5em{\scshape i\kern-0.25em b}\kern-0.8em\TeX}}}

\setcopyright{acmcopyright}
\copyrightyear{2018}
\acmYear{2018}
\acmDOI{XXXXXXX.XXXXXXX}

\acmConference[Conference acronym 'XX]{Make sure to enter the correct
  conference title from your rights confirmation email}{June 03--05,
  2018}{Woodstock, NY}
\acmPrice{15.00}
\acmISBN{978-1-4503-XXXX-X/18/06}

\newcommand{\subheading}[1]{\noindent{\textbf{#1}}}

\begin{document}

\title{Enhancing Malware Fingerprinting through Analysis of
Evasive Techniques}


\author{Alsharif Abuadbba}
\affiliation{%
  \institution{CSIRO's Data61}
  \country{Australia}}
\email{sharif.abuadbba@data61.csiro.au}

\author{Sean Lamont}
\affiliation{%
  \institution{Defence Science and Technology Group}
  \country{Australia}}
\email{sean.lamont2@defence.gov.au}

\author{Ejaz Ahmed}
\affiliation{%
  \institution{CSIRO's Data61}
  \country{Australia}}
\email{ejaz.ahmed@data61.csiro.au}

\author{Cody Christopher}
\affiliation{%
  \institution{CSIRO's Data61}
  \country{Australia}}
\email{cody.christopher@data61.csiro.au}

\author{Muhammad Ikram}
\affiliation{%
  \institution{Macquarie University}
  \country{Australia}}
\email{muhammad.ikram@mq.edu.au}

\author{Uday Tupakula}
\affiliation{%
  \institution{University of Newcastle}
  \country{Australia}}
\email{uday.tupakula@newcastle.edu.au}

\author{Daniel Coscia}
\affiliation{%
  \institution{Defence Science and Technology Group}
  \country{Australia}}
\email{daniel.coscia1@defence.gov.au}
\author{Mohamed Ali Kaafar}
\affiliation{%
  \institution{Macquarie University}
  \country{Australia}}
\email{dali.kaafar@mq.edu.au}
\author{Surya Nepal}
\affiliation{%
  \institution{CSIRO's Data61}
  \country{Australia}}
\email{surya.nepal@data61.csiro.au}


\begin{abstract}

As malware detection methods become more advanced and widespread, malware authors respond by adopting more sophisticated evasion mechanisms. However, traditional file-level fingerprinting, such as those provided by cryptographic and fuzzy hashes, is often overlooked as a potential target of evasion. Small variants within binary components are commonly used in malware attacks to evade traditional fingerprinting, as confirmed by Microsoft's discovery of the GoldMax variations in 2020 and 2021. Despite this, no large-scale empirical studies have investigated the limitations of traditional malware fingerprinting methods applied to actual samples obtained from the wild and how their effectiveness could be improved.

This paper addresses these gaps by answering three key questions: (a) To what extent are file variants commonly used in malware samples? To answer this question, we conduct an empirical study of a large-scale dataset of 4 million Windows Portable Executable (PE) files reports, 21 million sections, and 48 million resources, which we split chronologically into four groups to validate our analysis. Our findings suggest a high prevalence of similarities in deeper parts of PE files between 70\% to 80\%, including similar APIs in Import Libraries and common executable sections. (b) What file variant evasive methods can be observed? To answer this question, we cluster files with high similarities in their Import Libraries and common executable sections and labelled these clusters \textit{``Resilient fingerprints''} after validating their maliciousness ground truth through the Virustotal vendor labels. We then conduct a qualitative analysis across our four groups to identify the variant parts of the top resilient fingerprints. Our findings indicate that non-functional mutations - such as alterations in the number of sections, virtual size, virtual address, and section names - are being used extensively as primary evasive variant tactics. We also identify two executable sections of interest that are similar within each resilient fingerprint, which we call malicious sections (high entropy > 5) and camouflage sections (entropy = 0). (c) How can we use these identified characteristics to improve malware fingerprinting? To answer this question, we proposed two novel approaches that enhance malware fingerprinting and enable the identification of resilient fingerprints.  Using a large dataset of 4 million feeds from VirusTotal, our findings indicate a large potential improvement in fingerprinting, with more than 50\% of malware being identified compared to 20\% using traditional fingerprinting approaches.

\end{abstract}




\maketitle

\section{Introduction}

Advances in static and dynamic malware analysis have significantly improved the ability to detect and prevent malware attacks. Static analysis techniques~\cite{anderson2018ember,downing2021deepreflect} involve examining the characteristics of a file without executing it, whereas dynamic analysis techniques observe the behavior of the software as it runs in a virtual environment~\cite{mariconti2016mamadroid,jindal2019neurlux,wang2020you}.  Static analysis is notably faster than dynamic analysis. However, when dealing with large datasets, the process of extracting all relevant file characteristics, including obtaining disassembly using commercial tools like IDA Pro~\cite{IDAPro}, can still be time-consuming. As a result, malware fingerprinting is often used for detection as the first {\it triaging} mechanism~\cite{sikorski2012practical} to narrow the search space of files for further investigation. 


Malware fingerprinting is primarily based on file hashing. The output of hashing, which is commonly used and shared between cybersecurity teams, takes the form of checksums or unique identifiers for first triaging. There are two types of hash-based file-level fingerprinting techniques: cryptographic hashing and fuzzy hashing~\cite{oliver2021designing}. It is common practice to use the SHA-256 cryptographic hash to query a knowledge database like VirusTotal~\cite{virustotal} to determine whether a file is malicious or not. Security analysts use hashes as labels, which are shared with other analysts to help them identify malware. They are subsequently queried in online repositories to determine if the hash has been previously identified. 

While file-level hash fingerprinting is fast and useful, traditional cryptographic hashing poses a challenge in identifying similar malware. This is because cryptographic hashing  relies on concealing the correlation between the original entity and the hash value. Even one character change in the original entity, such as a file, results in a radically different hash, making identifying similar malware challenging. Another type of malware fingerprinting is fuzzy hashing. This overcomes the limitations of cryptographic hashing by producing similar hashes even when files are slightly changed, thereby making it more tolerant to evasion methods~\cite{tlsh}. Microsoft 365 recently confirmed that the GoldMax malware sample was first submitted to VirusTotal in September 2020, then reappeared in June 2021 with a new hash~\cite{goldmaxMS}. Despite some variation, both versions of GoldMax were captured by fuzzy hashing as the same, which cryptographic hashing failed to do. 

Security analysts, including Microsoft Threat Intelligence Center, have reported indications (e.g., GoldMax malware) that attackers have exploited the limitations of both types of hashes by creating multiple copies of sophisticated malware with variations resulting in significantly different new ones~\cite{goldmaxMS}. However, while there have been many studies investigating various malware-related activities, such as living off the land~\cite{barr2021survivalism}, IoT~\cite{alrawi2021circle}, Web~\cite{yao2023hiding}, Linux~\cite{cozzi2018understanding}, and sandboxing~\cite{yong2021inside}, there have been no systematic large-scale empirical studies investigating malware fingerprinting from real samples in the wild. Conducting such a study would answer the important question: Is there a large number of functionally similar malware being undetected by traditional fingerprinting methods? 
Strengthening the malware fingerprinting stage is crucial for the early triaging of potentially a large number of files, as it can enhance the explainability and effectiveness of subsequent, more extensive, and expensive forms of analysis. 

This motivates our study, which we address through the following three research questions (RQs):\\

{\bf RQ1.} To what extent are file variations commonly used in malware samples?

Building on this, we try to shed light on some of the trends in the current threat landscape to identify:

{\bf RQ2.}	What file variation evasive methods can be observed in each malware sample?

{\bf RQ3.}	How can we use the identified invariant file parts to improve malware fingerprinting? 

\begin{figure}[t]
    \centering
    \includegraphics[width=0.9\linewidth]{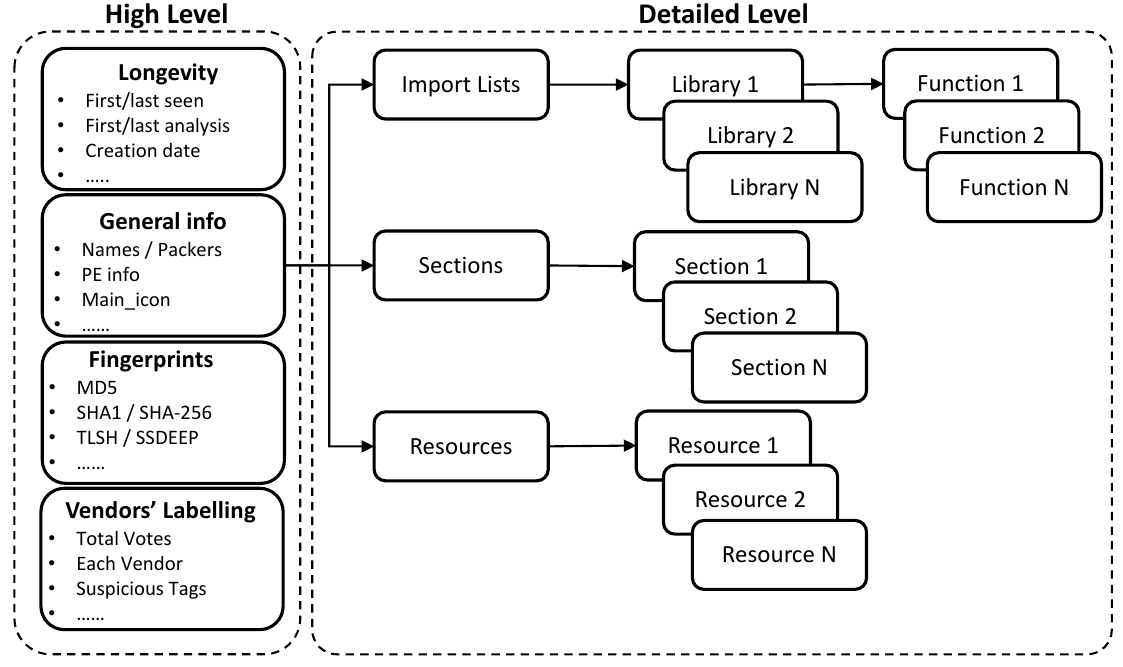}
    \caption{VirusTotal report granularity per file.}
    \label{fig:VT_highlevel}
    \vspace{-4mm}
\end{figure}

To study to what extent file variations are used within malware (RQ1), we conduct a systematic analysis of high-level file information and detailed metadata to investigate the prevalence of file variations. We collect a large-scale dataset from VirusTotal, which contains all reported files worldwide on a per-minute basis, over two separate periods of time (9 months apart), including detailed reports for each suspicious file at both high-level and detailed levels (see Section~\ref{sec:methodology} for further details). To ensure consistency in our analysis and findings, we focus on PE 
files of Windows operating systems, as they constitute up to 50\% 
all submitted files. 
We divide the dataset into four chronological groups to investigate the generalisability of our findings across different time periods. 
After examination, we found that traditional fingerprinting techniques have limited effectiveness in detecting similar malware files --- \textit{Cryptographic hash can only identify 16\% to 18.5\% of the files, while fuzzy hash performs slightly better, detecting 17.9\% to 20.7\% of the files}.
Analyzing deeper parts of the files as depicted in Fig.~\ref{fig:VT_highlevel} indicates a significant number of them shared common characteristics. Specifically, approximately 80\% of the files contained common Import Lists, while around 70\% of the files had some similar sections. Also, roughly 80\% of the files had comparable resources, as elaborated in Section~\ref{sec:prevelance}.

To address RQ2, we delve deeper into the files with common connections and investigate both the variant and invariant components of the malware samples. Firstly, we establish the ground truth of these identified files and ensure their malicious nature using the VirusTotal vendor labels. Secondly, we categorize the files with variations (different traditional hash) but common critical invariant components into groups called \textit{``Resilient fingerprints''} and perform a qualitative analysis.  Our findings suggest that two invariant components within the malware files could be used to connect and detect a deep fingerprint together. 
These invariant components are the Import Libraries, which contain a list of functionalities and common sections --- we refer to as ``camouflage sections'' and ``malicious sections'', as explained in Section~\ref{sec:malwarevariation}. We find the size of these resilient fingerprints can be quite large (e.g., 27,091 files). On the other hand, we also \textit{identify many non-functional variant evasive parts that are mutated between these files to obfuscate the hash, including the number of sections, virtual size, virtual address, and section names} (outlined in Section~\ref{sec:malwarevariation}).

To address RQ3, we aim to enhance malware fingerprinting by leveraging the findings from the previous two RQs. We propose two clustering approaches, Top-Down and Bottom-Up. As depicted in VirusTotal detailed level in Figure~\ref{fig:VT_highlevel}, the Top-Down approach initially clusters resilient fingerprints based on similarities in their Import Libraries and a list of functionalities. It then proceeds to identify common camouflage and malicious sections within that cluster. On the other hand, the Bottom-Up approach clusters all files from lower levels relying on common sections, such as camouflage and malicious sections. Our proposed methods have demonstrated consistent performance across four datasets. \textit{Both approaches have successfully identified between 30\% (Top-Down) and 58\% (Bottom-Up) of reported suspicious files connected using resilient fingerprinting strategies}, outperforming traditional fingerprinting mechanisms by up to three times. 
Our work is intended to augment the existing traditional fingerprinting techniques, rather than replace them altogether. Our proposed approaches are not to classify files as malicious or benign but rather  have been designed to enhance early triaging efforts by implementing a more comprehensive and thorough fingerprinting mechanism. This mechanism can capture a broader spectrum of variations of common malware files, as detailed in Section~\ref{sec:detectionimprovment}.

By answering the three RQs, this paper makes the following main contributions:

\begin{itemize}
    \item To the best of our knowledge, this paper presents the first large-scale systematic analysis investigating the prevalence of malware file variations and quantifying the limitations of traditional file-level fingerprinting.
    
    \item We systematically identify file variation evasive methods for bypassing traditional fingerprinting, and identify significant invariant components that can be used to improve early triage fingerprinting.
    
    \item We propose two novel approaches, Top-Down and Bottom-Up, that allow for identifying similar files at different levels, improving the efficacy of the early triage process.

    \item Our evaluation is extensive and based on a dataset of live feeds collected from VirusTotal, including over 4 million file reports, 21 million sections, and 48 million resources. Our results indicate that similarities within low-level parts are highly prevalent, while traditional hash-based methods are less effective, achieving a detection rate of only ~20\% within the 4 million feeds. Using the invariant parts we identified, we demonstrate that early triaging fingerprinting can be improved by more than ~50\%, representing a significant improvement over traditional methods. Furthermore, we plan to make all of our code and analysis public, providing researchers and practitioners with valuable resources to inform future research in the field.
\end{itemize}











\section{Background}\label{sec:background}

\begin{figure}[h]
    \centering
    \includegraphics[width=0.95\linewidth]{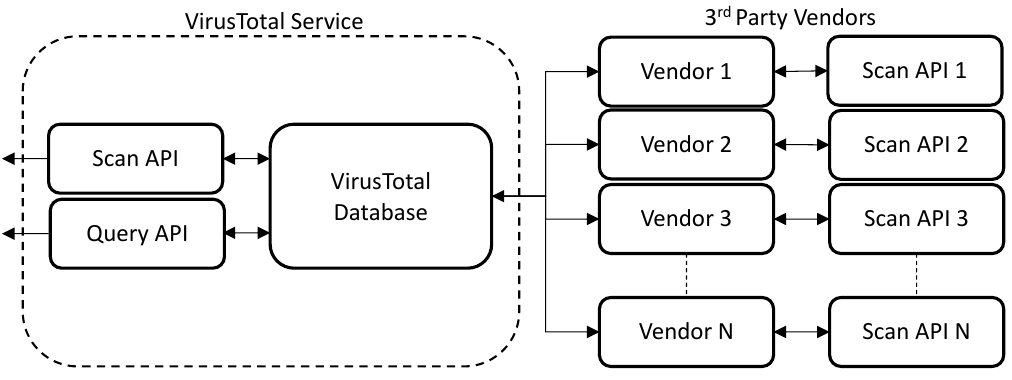}
    \caption{VirusTotal Feeds APIs and interaction with third-party security vendors overview.}
    \label{fig:Virutotal}
\end{figure}

\subheading{VirusTotal Platform.}
VirusTotal is a widely known service that scans files and web addresses to detect whether they are malicious among many existing techniques~\cite{virustotal,sabir2022reliability, shmalko2022profiler,mousavi2024investigation,abuadbba2022towards,evans2022raider}. In particular, file scanning is one of the most important types of features in the VirusTotal API and aims to detect files that deliver malware. As can be seen in Figure~\ref{fig:Virutotal}, VirusTotal collaborates with 71 external third-party security providers (see the full list at~\cite{virustotal}). Following the submission of a file to VirusTotal via the scanning API~\cite{vt_scanapi}, VirusTotal performs two actions:  (1) searches for an already existing file hash (SHA256) within their cached database, if the hash does not exist, then (2) forward the file to these vendors (i.e., anti-virus engines or online scanning services). 
VirusTotal retains records of the scanned files and corresponding results obtained from vendors in its database, which can be queried via the VirusTotal Report API~\cite{vt_scanapi}.

\subheading{Malware Fingerprinting and Vendor Labeling.}\label{sec:malware:labels}
The research community has extensively used VirusTotal to identify malicious files~\cite{invernizzi2014nazca,kolodenker2017paybreak,kotzias2016measuring,sharif2018predicting,sarabi2018characterizing,schwartz2018using,thomas2016investigating,wong2018tackling} as well as suspicious IP addresses and URLs~\cite{wang2014whowas,tian2018needle,wwwchain19,zhao2019decade}. Researchers usually integrate the labels from several vendors because vendors frequently disagree with one another as they are using different mechanisms to check those files. Remember that when a file is provided, VirusTotal returns 71 labels from the vendors. We find that the majority of papers specify a threshold $t$: the file is considered malicious if at least $t$ vendors return a ``malicious'' label. The majority of papers set $t$ = 2 to 4~\cite{invernizzi2014nazca,kolodenker2017paybreak,kotzias2016measuring,sharif2018predicting,wang2014whowas}, while a few papers further relax $t$ = 1~\cite{sarabi2018characterizing,schwartz2018using,thomas2016investigating,tian2018needle,wong2018tackling}. On the other hand, rarely researchers use a large threshold such as 40 when labelling malware files~\cite{cai2016inferring,korczynski2017capturing}, which has been shown to be ineffective~\cite{zhu2020measuring}. \textit{We therefore use a conservative $t$=4 following the majority of literature when deciding the ground truth of the collected samples}.

\section{Methodology}\label{sec:methodology}
We perform a systematic large-scale empirical study on real-world malware samples to examine how they vary evasive strategies. In this section, we detail our approach to data collection, data characteristics, and our systematic investigation of the aforementioned three research questions.

\subsection{Data Collection}
We utilize the VirusTotal API, specifically the File Feeds, which provide detailed, JSON-encoded reports about examined files worldwide, generated in batches every minute. Each batch is a text file containing one JSON structure per line, detailing the file's header, import list functions, sections, and resources, as depicted in Figure~\ref{fig:VT_highlevel}. These reports are automatically collected using a scheduled cron job. To validate our findings, we gathered VirusTotal batches from two periods: ten weeks from March 29th, 2022, to June 25th, 2022, and three weeks from March 27th, 2023, to April 17th, 2023.



\subsection{Data Characteristics}
\textbf{File Types.} We analyze the file types in the collected batches using the ``\textit{FileType}'' tag within each report, which is defined by the VirusTotal community. The top ten most common file types identified are Win32 EXE, HTML, Win32 DLL, Win64 EXE, XML, ZIP, TXT, Win64 DLL, ELF executable, and PDF. Approximately 50\% of the files are PE files from Windows operating systems, consistent with the findings of ~\cite{song2016learning}. PE files, including Win32 EXE, Win32 DLL, Win64 EXE, and Win64 DLL, are selected for investigating file variations due to their abundance and suitability for a reliable empirical study.


 


\subheading{PE File Report Structure.} The report structure for a file can be categorized into two parts -- high level and detailed level. The high level part, as shown in Figure~\ref{fig:VT_highlevel}, is primarily metadata such as submission date/time, type, hash (MD5, SHA256, and TLSH) and size. It also includes the vendors' votes about the maliciousness of the file. The detailed part of the report focuses on deeper segments of the files. This includes the import libraries, list functionalities, sections of code blocks and resources containing, for example, icons. For each of these, VirusTotal provides additional detailed metadata, such as the number of sections or imports. There are also detailed characteristics of each section, including its content hash, entropy, name, virtual size, and virtual address.


\subsection{Systematic Process}

To ensure study reliability, we divide the collected data from March to June 2022 into three groups (1-3), and the second dataset from March to April 2023 into the fourth group. Each group contains slightly over three weeks of file reports, with approximately 1,061,151 reports per group (Table~\ref{tb:group_details}). A relationship is observed between files and their number of sections and resources. On average, files have around 5 sections and 12 resources, with most having 3 sections and a smaller portion having a larger number of sections (>9). Additionally, Table~\ref{tb:filetype} presents the file types distribution. The majority of the files (around 60\%) are Win32 EXE, followed by Win32 DLL files (approximately 19\%). The combined percentage of Win64 EXE/DLL files is less than 20\%. Notably, a significant portion of the files (between 34\% and 42\%) are categorized as benign by VirusTotal vendor labeling (refer to Table~\ref{tb:benignLabels} in Appendix \ref{appendix:labels} for detailed label statistics).

\begin{table}[h]
\caption{Groups files, sections and resources detail.}
\vspace{-0.3cm}
\resizebox{2.5in}{!}{
\begin{tabular}{l|r|r|r|r}
\hline
          & {\bf Group 1} & {\bf Group 2} & {\bf Group 3} & {\bf Group 4}\\ \hline
Files     & 1,061,151  & 1,061,151  & 1,061,151  & 1,061,151\\ \hline
Sections  & 5,679,597  & 5,481,258  & 5,441,488  & 6,263,307\\ \hline
Resources & 12,589,924 & 12,715,782 & 11,159,391 & 13,627,706\\ \hline
\end{tabular}}
\label{tb:group_details}
\end{table}


\begin{table}[h]
\centering
\caption{File Types distribution of four groups.}
\vspace{-0.3cm}
\label{tb:filetype}
\resizebox{2.8in}{!}{
\begin{tabular}{c|c|c|c|c|c|c|c|c} 
\hline
          & \multicolumn{2}{c|}{Group 1} & \multicolumn{2}{c|}{Group 2} & \multicolumn{2}{c|}{Group 3} & \multicolumn{2}{c}{Group 4}  \\ 
\hline
File Type & Count   & \%                 & Count   & \%                 & Count   & \%                 & Count   & \%                  \\ 
\hline
Win32 EXE & 666,583 & 62.8               & 643,955 & 60.6               & 612,417 & 57.7               & 801,322 & 75.5                \\ 
\hline
Win32 DLL & 207,323 & 19.5               & 221,605 & 20.8               & 202,667 & 19.1               & 130,598 & 12.3                \\ 
\hline
Win64 EXE & 104,251 & 9.8                & 105,580 & 9.9                & 95,553  & 9.0                & 67,292  & 6.3                 \\ 
\hline
Win64 DLL & 82,899  & 7.8                & 80,012  & 7.5                & 77,285  & 7.2                & 61,844  & 5.8                 \\
\hline
\end{tabular}}
\end{table}



\section{Prevalence of File Variation}\label{sec:prevelance}
This section's focus is on answering RQ 1: \textit{To what extent are file variations commonly used in malware samples?} Here we aim to investigate the similarity of files at a deeper level than that given by traditional fingerprinting. 
We start with an overview, then focus on various file components where we believe similarities could be captured with examples. We finally present the obtained results and how consistent they are among the four groups. 

\subsection{Overview}
We start by examining how many similar files can be identified using traditional file fingerprinting techniques. We take SHA256 and TLSH to represent cryptographic and fuzzy hashing respectively. 

The obtained results seem consistent across the 4 groups with low similarities of 16\%-20\% (see Results Section~\ref{sec:results}). We then investigate the prevalence of non-functional file variations, which are done to alter the hash of the file. This is an evasion technique employed by malware authors to bypass traditional hash-based fingerprinting. For example, a malware author may modify a previously detected malicious file by padding the file with non-functional content, thereby changing the hash and avoiding detection. We identify three parts within a file that are representative of the actual intent of the file. When these are common between two files with unique hashes, this is indicative of file variations. These parts are the import libraries' list of functionalities, sections, and resources. Next, we elaborate further on each of these parts with illustrations before presenting the obtained results. 

\subsection{Import Libraries with List of Functions}\label{sec:importList}
\begin{figure}[h]
    \centering
    \includegraphics[width=0.8\linewidth]{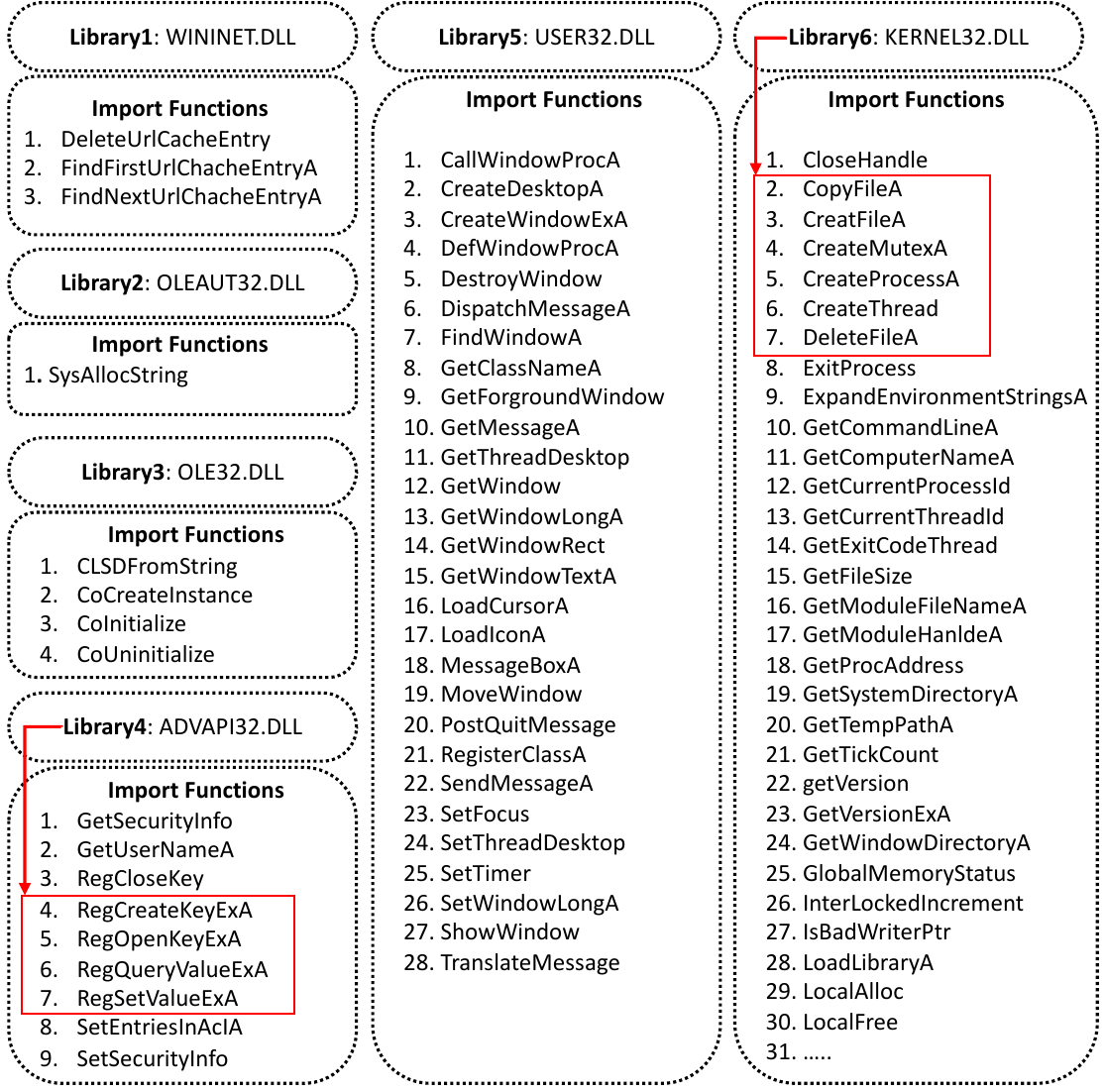}
    \caption{Import List -- each pair of boxes shows the imported library and a list of functions from that library.}
    \label{fig:importlist}
\end{figure}

The import list contains a detailed list of functions imported from other libraries that are used by  files to achieve their intent. 
Figure~\ref{fig:importlist} is an example import list extracted from a sample malware file. The file imports six libraries and a few functions from each library. Observing these functions, security analysts would typically infer some of the file intentions. For example, the file is importing (Library 4: \textit{ADVAPI32.DLL}), which is known to provide access to core Windows components such as Service Manager and Registry. 
Observing functions 4-7 from this library, we can see that the malware file actually alters the Registry by creating an entry and setting up its value. The second example is the import (Library 6: \textit{KERNEL32.DLL}), which provides core functionalities such as access and manipulation of files, memory, and hardware. Looking at some of the highlighted functions 2-7, we observe the file actually copies files, creates processes, and then deletes those files. 

It is clear that observing the similarities of the import list would be a reliable indication of common intent. We adopt a cautious approach by stipulating that for us to consider two files as variations, they must exhibit a 100\% similarity of the entire imported list of libraries and imported functions while having different file hashes. 

\subsection{Sections of PE Files - Executable Code}
A PE file comprises several sections, each containing executable code or relevant details related to the executable code. The sections could be one of the following types: \textit{``.text''}  contains the executable code; \textit{``.rdata''} holds read-only data that is globally accessible within the program; \textit{``.data''} stores global data accessed throughout the program; \textit{``.rsrc''} stores resources needed by the executable.

Having two files sharing the same code blocks while producing two different file hashes might be an indication of malware file variation. Consequently, we use sections as another indication of the prevalence of file variations. Through the collected VirusTotal feeds, we have detailed information about every section within the file. For example, Table~\ref{fig:sectionData} shows one of four sections of a real malware sample we collected. The table shows that the sections' metadata has fine-grained details, including its content hash, name/type, entropy, raw\_size, virtual\_address, and virtual\_size, which could be a reliable source to identify the similarities.  


\begin{table}
\centering
\caption{Section metadata provided by VirusTotal.}
\vspace{-0.3cm}
\label{fig:sectionData}
\resizebox{2in}{!}{
\begin{tabular}{r|l} 
\hline
{\bf Metadata}         & {\bf Value}                             \\ 
\hline
MD5              & a0a2aedaaf49ad428951a94bf6038890  \\ 
\hline
Chi-squared      & 80,784.76                          \\ 
\hline
Entropy          & 7.5                               \\ 
\hline
Name             & updateRegistry.text                             \\ 
\hline
Raw\_size        & 32,340                             \\ 
\hline
Virtual\_Address & 4,096                              \\ 
\hline
Virtual\_Size    & 32,340                             \\ 
\hline
Flags    & {\tt rwx}                             \\ 
\hline
\end{tabular}}
\vspace{-3mm}
\end{table}
\subsection{Resources of PE Files -- non-Executables}
The resources part of a file contains non-executable materials such as icons, images and strings. Resource types could include Data, ASCII text, Image/x-png, Lotus 1-2-3, Image/gif, or Audio/mpeg. Our findings suggest that resources may be less effective in indicating the intent of a file as they can be similar between behaviorally different files. This is because these resources, such as Microsoft logos, may be used by both benign and malware samples (refer to Appendix~\ref{sec:resourcesUnrialiable} for further detail). Similar to sections, the available resource's metadata from VirusTotal feeds has detailed information, including its content hash, type, entropy, chi2, and typeX.  


\subsection{Results}\label{sec:results}
\begin{figure}[h]
    \centering
    \includegraphics[width=0.8\linewidth]{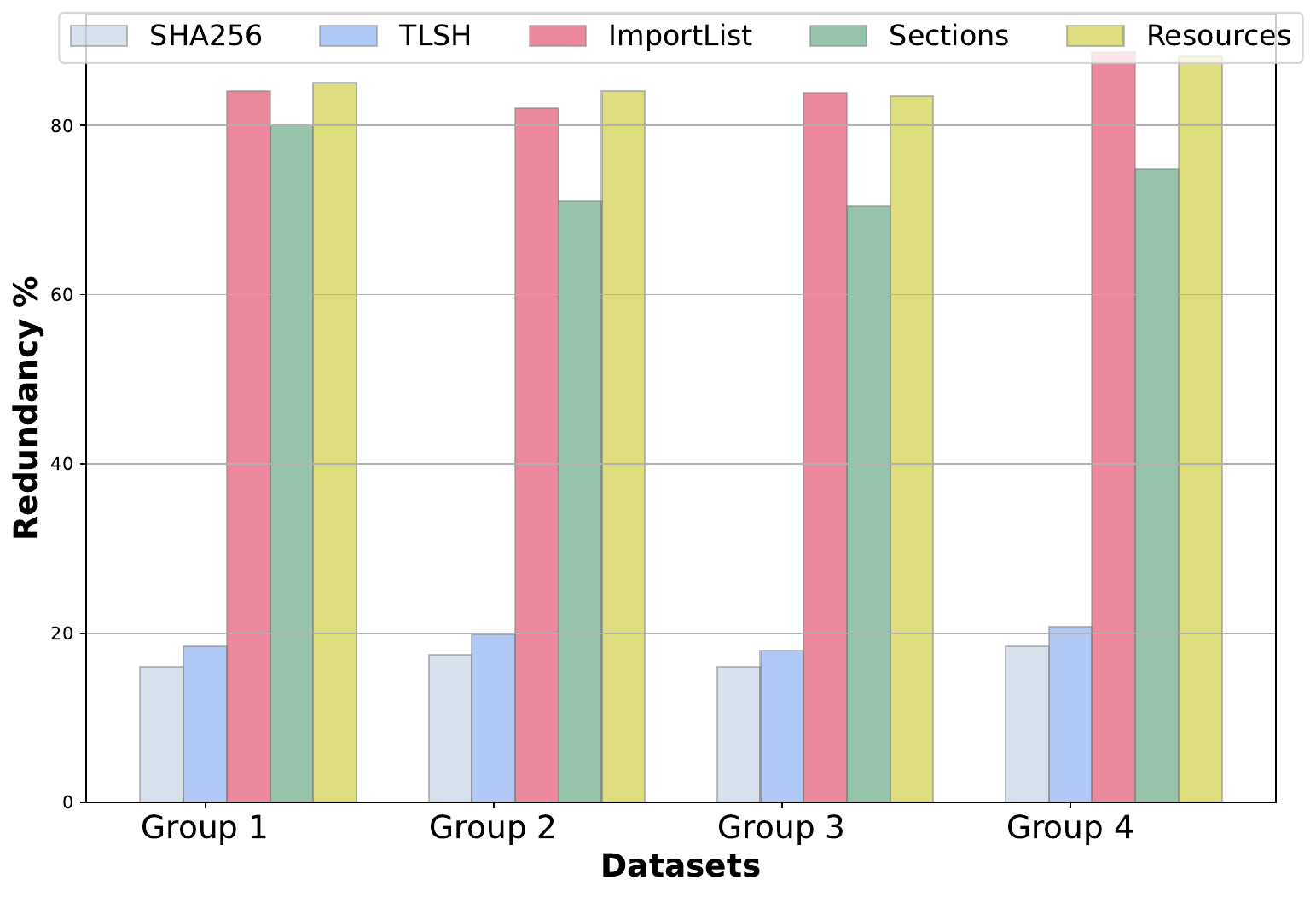}
     \vspace{-0.2cm}
    \caption{File Prevalence.}
    \label{fig:fileprevalnce}
    \vspace{-4mm}
\end{figure}

We cluster based on content similarity match using the above three critical file parts to compare against the traditional file level hash fingerprinting. Figure~\ref{fig:fileprevalnce} shows the obtained results across the four groups. To calculate redundancy on the y-axis, we follow these steps. Firstly, we apply a unique function to a set of $N$ files hashes (e.g., 100). This process helps us determine the count of unique hashes, represented as $U$ (e.g., 84). To find the number of redundant hashes, we subtract $U$ from $N$, resulting in 16 (100 - 84 = 16).  It is clear that traditional file level hash (SHA256 and TLSH) could only identify between 16\%-20\% file redundancy within each group.  However, when we dig deeper into various critical parts of the files like the import list, sections, and resources, we could see the similarity of 82\%-84\% using the import list of libraries and used functions, 70\%-80\% using the part of the sections, and 83\%-85\% using the resources. These findings have motivated us to delve deeper into identifying the specific variant parts and to investigate how we can leverage these similarities at a more profound level to develop sophisticated fingerprinting techniques. 


\observ{\small Recent trends in the malware field indicate a widespread adoption of file variation. Analyzing over 4 million files, 21 million sections, and 48 million resources, the findings reveal that traditional file hash identifies only <20\% redundancy. However, delving further into file components uncovers over 80\% similarities, making it a valuable initial triaging mechanism}
\section{Malware Variation Evasive Methods}\label{sec:malwarevariation}

In this section, we aim to address RQ 2: \textit{What file variation evasive methods can we observe per malware sample?} To accomplish this, we establish two key elements. We first develop a method for grouping different malware files with the same behaviour into clusters. We refer to these as \textit{``Resilient fingerprints''}, which we define formally in Section~\ref{sec:campaginDef}. Variations within these groups answer the question of what evasive methods are used. Secondly, we confirm that these clusters are indeed malicious, as discussed in Section~\ref{sec:campAnalysis}. Finally, we will delve deeper into the identified resilient fingerprints to identify both variant and invariant parts, as presented in Sections~\ref{sec:variation} and \ref{sec:invariation}. We also identify certain unreliable invariant components (see Appendix \ref{sec:resourcesUnrialiable}).

\subsection{Resilient Fingerprints Definition}\label{sec:campaginDef}
In this section, we introduce the concept of ``\textit{Resilient fingerprints}'' for the purpose of our analysis. A resilient fingerprint refers to a set of PE files that share the same system functionalities while appearing different at a surface level (i.e., having different file-level hashes). We have found that the most informative aspect of a file that captures its intent is the set of import libraries and the functions they contain, as well as their executable sections, as described in Section~\ref{sec:importList} and illustrated in Figure~\ref{fig:importlist}. These libraries and functions reflect the system functionalities that the file calls upon to perform its actions. For example, the presence of the \textit{Advapi32.dll} library with functions such as \textit{RegCreateKeyExA, RegOpenKeyExA, RegQueryValueExA, and RegSetValueExA} in a file's import libraries would indicate tampering with the Registry. Similarly, the presence of the \textit{KERNEL32.dll} library with functions such as \textit{CopyFileA, CreateFileA, CreateProcessA, CreateThread, and DeleteFileA} would indicate memory and file operations.
Our analysis considers two files to be part of the same resilient fingerprint if they have 100\% similar import libraries and functions, followed by an examination of common executable sections. While these thresholds can be adjusted with certain percentages, we find that files must have common executable section code blocks in addition to the import lists to execute that intent, as determined through thorough analysis.


\vspace{-0.1cm}
\subsection{Resilient Fingerprints Qualitative Analysis}\label{sec:campAnalysis}

In this paper, we thoroughly investigate the resilient fingerprints within the four groups' datasets by examining their ground truth labels using data collected from VirusTotal vendors. Our focus in this section is on a qualitative analysis of the top 4-5 resilient fingerprints from each group with a significant number of files (threshold >= 900). To gain a deeper understanding of each resilient fingerprint, we analyze its intent, tag (a simplified version of a unique identifier), type (the operating system file targeted), characteristics (malicious activities), file number (number of files identified with similar import lists and functionalities), longevity (timestamp of the campaign), and file level hash fingerprint (the percentage of identified files with similar file hashes that have been detected using traditional fingerprinting methods such as SHA256).

\subheading{Findings.} We summarize our findings in Table~\ref{tb:topcampaigns}, identifying 18 resilient fingerprints with intents like Worm Autospread, Trojan Zombie, Trojan Spyware, Trojan Mira, and Trojan-Dropper. Noteworthy insights include:
(1) These fingerprints have large sizes, up to 34,068 files, but traditional SHA256 file-level hash fingerprinting methods perform poorly (15 out of 18 cases <10\%) in detecting them as similar due to evasion techniques employed by malware authors.
(2) They target both Win32/64 and EXE/DLL, indicating widespread use of evasion tactics.
(3) Some fingerprints date back to 1992 and continue until 2022, highlighting the presence of long-standing malware utilizing known system functionalities that go undetected by traditional fingerprinting mechanisms.
(4) Certain resilient fingerprints reappear in multiple dataset groups, such as Worm Autospread with ID 35b0 (marked in pink) and Trojan Mira with ID 3a58 (marked in green), consistently ranking among the largest identified resilient fingerprints in most of the four groups. Additionally, Trojan Spyware with ID 25ca (marked in orange) and Trojan Spyware with ID 834c (marked in purple), appear across the first/second and second/third groups.

\subheading{Maliciousness Ground Truth.} 
Before investigating the variant evasive and invariant parts within the files of an identified campaign, it is necessary to establish the ground truth of maliciousness through other means, as traditional hash fingerprinting methods have a low similarity detection score. In this study, we leverage the vendors' labels within the collected groups' dataset to establish this ground truth.
We determine the number of vendors that labeled each file as malicious, $x$, and construct a vector per resilient fingerprint, $r={x_1,x_2,...,x_n}$, where $x_1$ represents file 1, $x_2$ represents file 2, and so on. We then calculate the histogram distribution of $r$. Figure~\ref{fig:groundTruthG} shows the maliciousness of ground truth labelling from VirusTotal vendors to the top 18 identified resilient fingerprints. \textit{Our findings confirm that all the files within each resilient fingerprint are highly malicious, as more than 20 vendors have labelled them as malicious} (refer to Figure~\ref{fig:groundTruthG} in Appendix \ref{appendix:18fingerprints} for details).

\begin{table}
\centering
\caption{Top  resilient fingerprints from the four groups.}
\label{tb:topcampaigns}
\vspace{-0.3cm}
\resizebox{3.5in}{!}{
\begin{tabular}{c|c|c|c|c|c|c} 
\hline
\multicolumn{7}{c}{\textbf{Group 1}}                                                                                                                                                                                                                                                                                                                                                                                                                                                                         \\ 
\hline
\rowcolor[rgb]{0.937,0.937,0.937} \textbf{Intent}                                                                                            & \textbf{ID}                                                   & \textbf{Type} & \textbf{Characteristics}                                                            & \textbf{Files No}           & \textbf{Longevity}             & \begin{tabular}[c]{@{}>{\cellcolor[rgb]{0.937,0.937,0.937}}c@{}}\textbf{File Hash }\\\textbf{ Fingerprint}\end{tabular}  \\ 
\hline
{\cellcolor[rgb]{1,0.8,0.788}}\begin{tabular}[c]{@{}>{\cellcolor[rgb]{1,0.8,0.788}}c@{}}Worm \\ Autospread\end{tabular}                      & \multicolumn{1}{l|}{{\cellcolor[rgb]{1,0.8,0.788}}35b0}       & Win32.exe     & \begin{tabular}[c]{@{}c@{}}Spread through \\ removables like USB\end{tabular}       & 27,091                      & 1999-2022                      & 3.7\%                                                                                                                    \\ 
\hline
\begin{tabular}[c]{@{}c@{}}Trojan \\ Zombie\end{tabular}                                                                                     & \multicolumn{1}{l|}{8a9e}                                     & Win32.exe     & \begin{tabular}[c]{@{}c@{}}Command \\ Control for attacks\end{tabular}              & 24,119                      & 2011-2022                      & 9.7\%                                                                                                                    \\ 
\hline
{\cellcolor[rgb]{1,0.8,0.404}}\begin{tabular}[c]{@{}>{\cellcolor[rgb]{1,0.8,0.404}}c@{}}Trojan \\ Spyware\end{tabular}                       & \multicolumn{1}{l|}{{\cellcolor[rgb]{1,0.8,0.404}}25ca}       & Win64.dll     & \begin{tabular}[c]{@{}c@{}}Stealing Banking \\ Credentials\end{tabular}             & 10,295                      & 2015-2022                      & 0.4\%                                                                                                                    \\ 
\hline
{\cellcolor[rgb]{0.769,0.886,0.769}}\begin{tabular}[c]{@{}>{\cellcolor[rgb]{0.769,0.886,0.769}}c@{}}Trojan \\ Mira\end{tabular}              & \multicolumn{1}{l|}{{\cellcolor[rgb]{0.769,0.886,0.769}}3a58} & Win32.dll     & \begin{tabular}[c]{@{}c@{}}Command Control\\ for attacks\end{tabular}               & 3,596                       & 2014-2022                      & 2.6\%                                                                                                                    \\ 
\hline
\begin{tabular}[c]{@{}c@{}}Trojan \\ Dropper\end{tabular}                                                                                    & \multicolumn{1}{l|}{fbc6}                                     & Win64.dll     & Install other Malware                                                               & 917                         & 2020-2022                      & 4.9\%                                                                                                                    \\ 
\hline
\multicolumn{7}{c}{\textbf{Group 2}}                                                                                                                                                                                                                                                                                                                                                                                                                                                                         \\ 
\hline
{\cellcolor[rgb]{1,0.8,0.788}}\begin{tabular}[c]{@{}>{\cellcolor[rgb]{1,0.8,0.788}}c@{}}Worm \\ Autospread\end{tabular}                      & \multicolumn{1}{l|}{{\cellcolor[rgb]{1,0.8,0.788}}35b0}       & Win32.exe     & \begin{tabular}[c]{@{}c@{}}Spread through \\ removables like USB\end{tabular}       & \multicolumn{1}{l|}{34,045} & \multicolumn{1}{l|}{1999-2022} & 7.4\%                                                                                                                    \\ 
\hline
Trojan Ransomeware                                                                                                                           & d66e                                                          & Win32.exe     & Encrypt files.                                                                      & 31,747                      & 2008-2022                      & 77.7\%                                                                                                                   \\ 
\hline
{\cellcolor[rgb]{0.796,0.808,0.984}}Trojan Spyware                                                                                           & {\cellcolor[rgb]{0.796,0.808,0.984}}834c                      & Win64.dll     & Keylogger.                                                                          & 17,379                      & 1992-2022                      & 3.0\%                                                                                                                    \\ 
\hline
{\cellcolor[rgb]{0.769,0.886,0.769}}Trojan Mira                                                                                              & {\cellcolor[rgb]{0.769,0.886,0.769}}3a58                      & Win32.dll     & \begin{tabular}[c]{@{}c@{}}Remote Command \\ Control\end{tabular}                   & 10,794                      & 2014-2022                      & 3.1\%                                                                                                                    \\ 
\hline
{\cellcolor[rgb]{1,0.8,0.404}}Trojan Spyware                                                                                                 & {\cellcolor[rgb]{1,0.8,0.404}}25ca                            & Win64.dll     & \begin{tabular}[c]{@{}c@{}}Stealing banking \\ credentials.\end{tabular}            & 2,070                       & 2015-2022                      & 5.2\%                                                                                                                    \\ 
\hline
\multicolumn{7}{c}{\textbf{Group 3}}                                                                                                                                                                                                                                                                                                                                                                                                                                                                         \\ 
\hline
{\cellcolor[rgb]{1,0.8,0.788}}Worm Autospread                                                                                                & {\cellcolor[rgb]{1,0.8,0.788}}35b0                            & Win32.exe     & \begin{tabular}[c]{@{}c@{}}Spread through \\ removables like USB\end{tabular}       & 34,068                      & 1992-2022                      & 5.6\%                                                                                                                    \\ 
\hline
Virus Fujaks                                                                                                                                 & 9dab                                                          & Win32.exe     & \begin{tabular}[c]{@{}c@{}}Prevent security \\ processes from running.\end{tabular} & 20,844                      & 1992-2022                      & 3.0\%                                                                                                                    \\ 
\hline
{\cellcolor[rgb]{0.796,0.808,0.984}}Trojan Spyware                                                                                           & {\cellcolor[rgb]{0.796,0.808,0.984}}834c                      & Win64.dll     & Keylogger.                                                                          & 18,068                      & 1992-2022                      & 2.8\%                                                                                                                    \\ 
\hline
{\cellcolor[rgb]{0.769,0.886,0.769}}Trojan Mira                                                                                              & {\cellcolor[rgb]{0.769,0.886,0.769}}3a58                      & Win32.dll     & \begin{tabular}[c]{@{}c@{}}Remote Command \\ Control\end{tabular}                   & 10,232                      & 2014-2022                      & 1.8\%                                                                                                                    \\ 
\hline
\multicolumn{7}{c}{\textbf{\textbf{Group 4 (Collected 9 months after)}}}                                                                                                                                                                                                                                                                                                                                                                                                                                     \\ 
\hline
\multicolumn{1}{c|}{Trojan Ransomeware}                                                                                                     & 5a7b                                                          & Win32.exe     & Encrypt Files                                                                       & 11,926                      & 2008-2022                      & \multicolumn{1}{c}{18.1\%}                                                                                              \\ 
\hline
\multicolumn{1}{c|}{Trojan Malware}                                                                                                         & 6448                                                          & Win32.exe     & \begin{tabular}[c]{@{}c@{}}Remote Command \\Control\end{tabular}                    & 8,842                       & 1992-2022                      & \multicolumn{1}{c}{37.3\%}                                                                                              \\ 
\hline
\multicolumn{1}{c|}{{\cellcolor[rgb]{1,0.8,0.788}}\begin{tabular}[c]{@{}>{\cellcolor[rgb]{1,0.8,0.788}}c@{}}Worm \\Autospread\end{tabular}} & {\cellcolor[rgb]{1,0.8,0.788}}35b0                            & Win32.exe     & \begin{tabular}[c]{@{}c@{}}Spread through \\~removables likeUSB~ ~ ~~\end{tabular}  & 1,979                       & 1992-2022                      & \multicolumn{1}{c}{6.7\%}                                                                                               \\ 
\hline
\multicolumn{1}{c|}{{\cellcolor[rgb]{0.769,0.886,0.769}}Trojan Mira}                                                                        & {\cellcolor[rgb]{0.769,0.886,0.769}}3a58                      & Win32.dll     & \begin{tabular}[c]{@{}c@{}}Remote Command\\Control\end{tabular}                     & 1,969                       & 2003-2022                      & \multicolumn{1}{c}{10.6\%}                                                                                              \\
\hline
\end{tabular}}
\end{table}

\subsection{Reliable Invariant  Executable Parts}\label{sec:invariation}

\subheading{Findings.} 
After thoroughly analysing the files' critical components, we discover that the identified resilient fingerprints utilize common executable section blocks. Specifically, we identify three types of invariant sections, and named them ``Malicious'', ``Standard'', and ``Camouflage''. The ``Malicious'' section is where the majority of malicious code is embedded, characterized by high entropy (>5) and frequent obfuscation. The ``Standard'' section appears to contain generic standard operation code blocks that may be called by the ``Malicious'' section and has normal entropy ($\leq5$). The ``Camouflage'' section is the most intriguing aspect we identified. It appears to be a padding section with an entropy of 0 and a small raw size (<4096 bytes), indicating the presence of automatically generated continuous characters. These findings were prevalent across all four groups.\\ 

\subheading{Results Insight 1.} Table~\ref{tb:invariationSec} provides a detailed examination of the invariant sections for the resilient fingerprint ``Worm Autospread'' (ID 35b0), which was observed across all four groups. This resilient fingerprint consists of five sections, each identified by a SecID, which is a shortened version of the first and last two characters of their content hash. The table also includes the entropy and count of each section within the fingerprint files for each group, as well as our analysis label. It is evident that these sections are present not only within the resilient fingerprint for each group dataset, but also across multiple groups. Out of the five sections, we can see that this resilient fingerprint includes two likely ``Malicious'', two ``Standard'', and one ``Camouflage'' sections.

\begin{table}[ht]
\caption{In-variant Sections. Zoom in to Worm Autospread resilient fingerprint with ID 35b0.}
\label{tb:invariationSec}
\vspace{-0.3cm}
\centering
\resizebox{2.8in}{!}{
\begin{tabular}{c|c|c|c|c|c|c}
\hline
{\bf SecID} & {\bf Entropy} & \begin{tabular}[c]{@{}c@{}}{\bf Count}\\ {\bf Group 1}\end{tabular} & \begin{tabular}[c]{@{}c@{}}{\bf Count}\\ {\bf Group 2}\end{tabular} & \begin{tabular}[c]{@{}c@{}}{\bf Count}\\ {\bf Group 3}\end{tabular} & \begin{tabular}[c]{@{}c@{}}{\bf Count}\\ {\bf Group 4}\end{tabular} & {\bf Label}      \\ \hline
d47e  & 0       & 27103                                                   & 34045                                                   & 34262                                                &2179   & Camouflage \\ \hline
31d3  & 7.89    & 15411                                                   & 16158                                                   & 17313                                                 &--   & Malicious  \\ \hline
80b1  & 2.81    & 15411                                                   & 16158                                                   & 17313                                                 &986   & Standard   \\ \hline
40af  & 7.91    & 11692                                                   & 17887                                                   & 16949                                                 &1193   & Malicious  \\ \hline
3a56  & 2.79    & 11692                                                   & 17887                                                   & 16949                                                 &1193   & Standard   \\ \hline
\end{tabular}}
\end{table}

\subheading{Results Insight 2.} Table~\ref{tb:invariationSec2} examines the invariant sections of another resilient fingerprint, ``Trojan Spyware'' (ID 25ca), which was observed twice across groups 1 and 2 of our datasets. In addition to the patterns discussed in Insight 1 regarding the three common sections, this fingerprint reveals that the malware authors also extensively utilize camouflage sections. For instance, four of the ten sections are camouflage, characterized by an entropy of 0 and simple padding of <4096 bytes. Additionally, the malware authors repeatedly include the same camouflage section within the same malware sample. For instance, SecID ``d47e'' appeared 157,453 times across 10,295 files or around 12-15 copies per file. These tactics, beyond altering the file hashes, are likely used also to increase the size of the malicious files, making it more challenging for the next stage of static and dynamic analysis.

\begin{table}[ht]
\caption{Invariant Sections. Zoom in to Trojan Spyware resilient fingerprint with ID 25ca and have 10295 files.}
\label{tb:invariationSec2}
\vspace{-0.3cm}
\resizebox{2.2in}{!}{
\begin{tabular}{c|c|c|c|c}
\hline
{\bf SecID} & {\bf Entropy} & \begin{tabular}[c]{@{}c@{}}{\bf Count}\\ {\bf Group 1}\end{tabular} & \begin{tabular}[c]{@{}c@{}}{\bf Count}\\ {\bf Group 2}\end{tabular} & {\bf Label}      \\ \hline
6210  & 0       & 157453                                                  & 33183                                                   & Camouflage \\ \hline
0822  & 0       & 83165                                                   & 17294                                                   & Camouflage \\ \hline
459b  & 0       & 12326                                                   & 2584                                                    & Camouflage \\ \hline
b66c  & 5.14    & 10295                                                   & 2070                                                    & Malicious  \\ \hline
a214  & 2.78    & 10295                                                   & 2070                                                    & Standard   \\ \hline
ca9d  & 4.04    & 10295                                                   & 2070                                                    & Standard   \\ \hline
bad4  & 0       & 10295                                                   & 2070                                                    & Camouflage \\ \hline
0e5c  & 4.09    & 10295                                                   & 2070                                                    & Standard   \\ \hline
07aa  & 7.82    & 10290                                                   & 2069                                                    & Malicious  \\ \hline
21ae  & 6.67    & 10290                                                   & 2069                                                    & Malicious  \\ \hline
\end{tabular}}
\end{table}

\begin{figure}[h]
    \centering
    \includegraphics[width=0.9\linewidth]{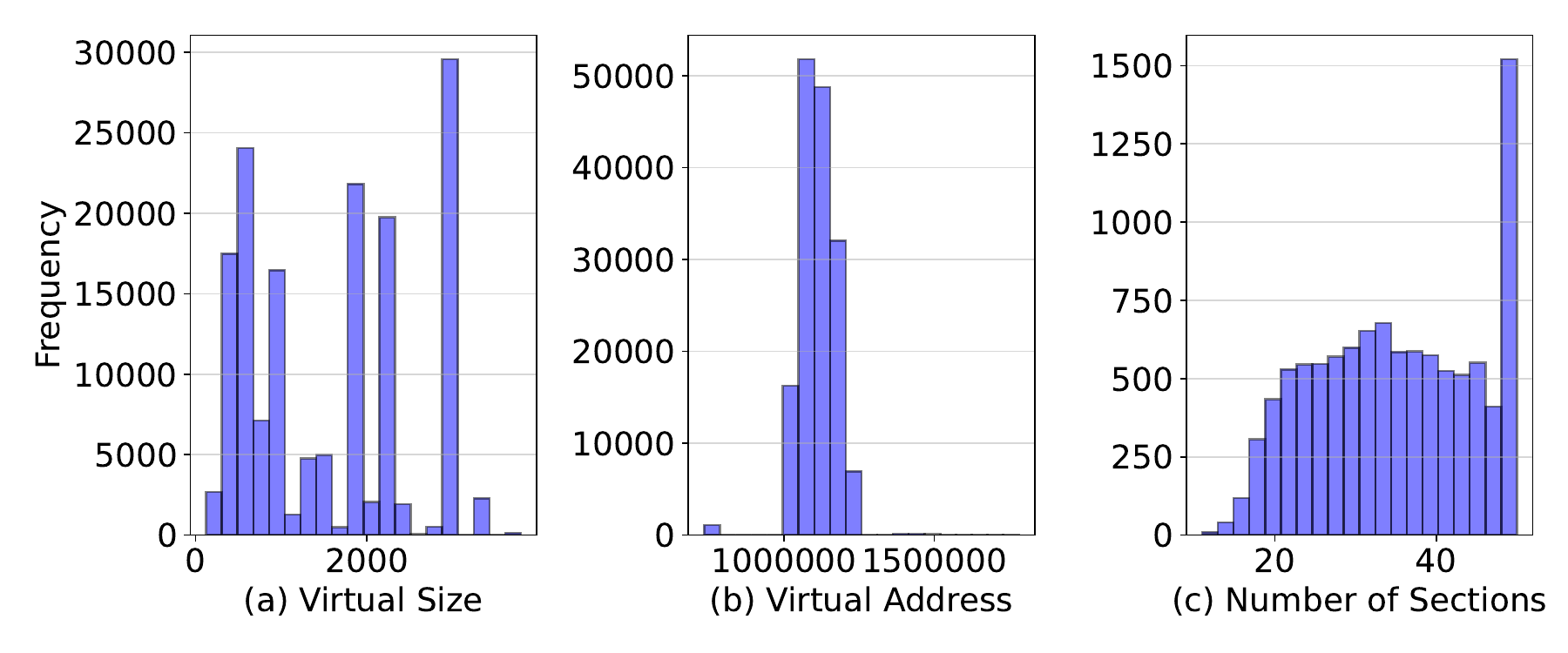}
     \vspace{-0.3cm}
    \caption{Section Numbers, Virtual Size and Address variant.}
    \label{fig:variantSize_sections}
    \vspace{-4mm}
\end{figure}

\begin{figure}[h]
    \centering
    \includegraphics[width=0.45\linewidth]{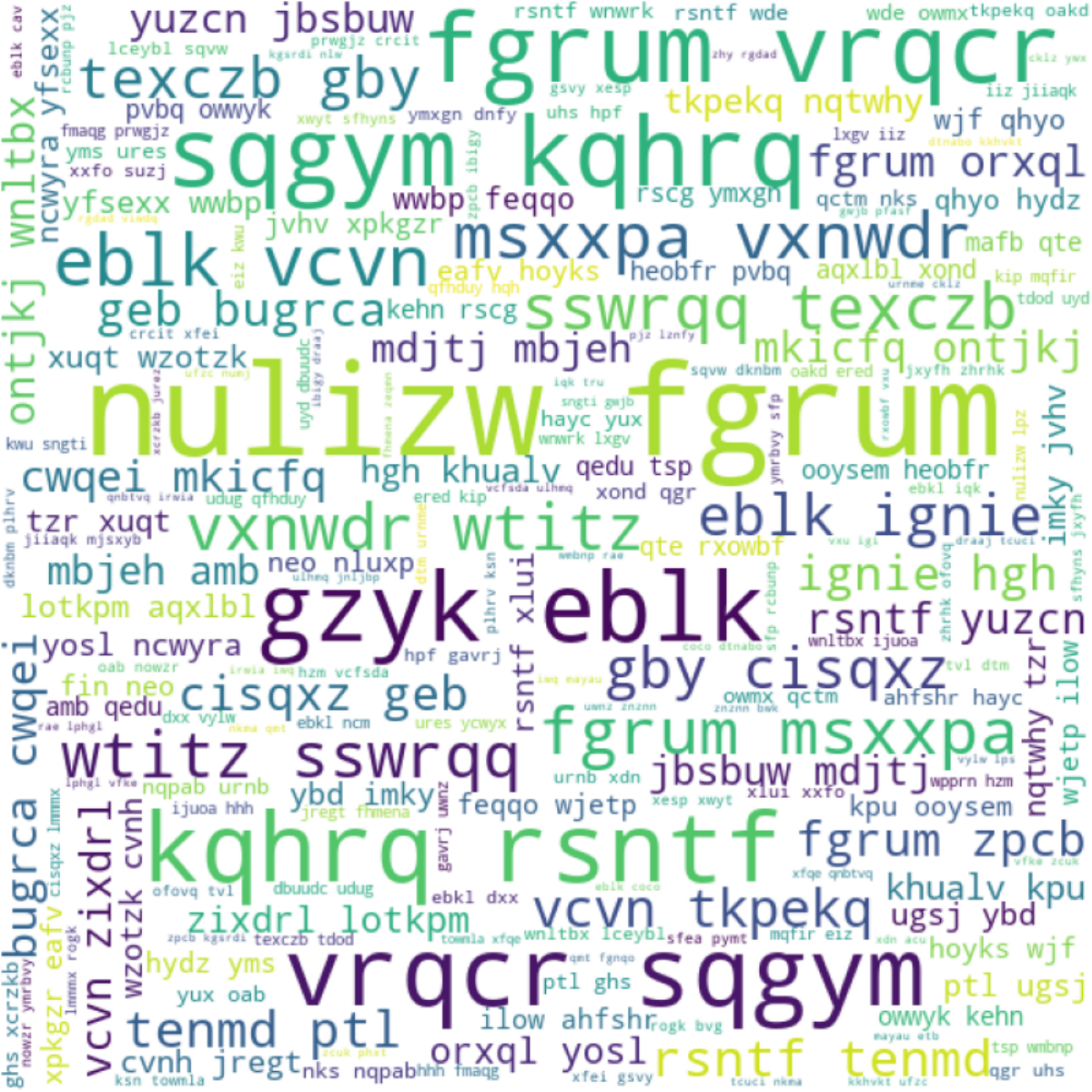}
     \vspace{0.2cm}
    \caption{Same section content across 10,295 files with variant section names.}
    \label{fig:secion_names}
    \vspace{-4mm}
\end{figure}

\subsection{Variant Evasive Parts}\label{sec:variation}
\subheading{Findings.} Our analysis of resilient fingerprint files reveals various tactics employed by malware authors to evade detection and hide their file hash fingerprints. These tactics focus on section numbers, virtual size, virtual address, and section names. (1) Malware authors frequently use high section numbers to inject camouflage sections and increase file size, making analysis more challenging. (2) Manipulating virtual size by utilizing uninitialized arrays in specific sections alters the memory size footprint. (3) The virtual address, the location in memory where the file is loaded, is also manipulated by loading the same sections into random memory locations. (4) Additionally, malware authors use variant section names to conceal common sections with identical content. Subsequent presentation provides evidence and results supporting these findings.


\subheading{Results Insight 1.} As shown in Figure~\ref{fig:variantSize_sections}. (c), the distribution of section numbers within the Trojan Spyware (ID 25ca) resilient fingerprint, which comprises 10,295 files and over 300,000 sections, reveals significant variability. Despite the presence of invariant section blocks, as previously discussed, the number of sections per file varies considerably, ranging from 10 to 50. A large proportion of these sections are attributed to camouflage sections (SecID 6210 and 0822, as shown in Table~\ref{tb:invariationSec2}), which are mostly empty sections.


\subheading{Results Insight 2.}  As Figure~\ref{fig:variantSize_sections}. (a) and (b)  illustrates, virtual size is highly variable among resilient fingerprint files, despite the presence of identical section content. This variability ranges from 0 to 3500. Similarly, the virtual address is widely dispersed, ranging from 0.1e6 to 1.6e6. Additionally, Figure~\ref{fig:secion_names} exhibits that the section names (with identical content) display high randomness, suggesting that they were auto-generated as a camouflage tactic.\\

It is important to note that the degree of variability across the variant parts we have discussed varies among fingerprints. Some resilient fingerprints may exhibit a higher degree of randomness in certain areas, such as section numbers, virtual addresses, virtual sizes, or section names, while others may exhibit less variability in these areas. To aid the reviewers further, we have made all identified top resilient fingerprints available in our repository\footnote{ ~\url{https://github.com/anonymousConfer/malware_fingerprint}}.

\observ{\small Import libraries with lists of functionalities and executable section blocks are reliable components for identifying invariant parts of malware samples through resilient fingerprints. Evasive methods used by malware include altering the number of sections, virtual size, virtual address, section names, and occasionally code size variations}
\section{Fingerprint detection Improvement}\label{sec:detectionimprovment}
This section focuses on answering RQ3: \textit{How to improve malware fingerprinting?} To achieve this, we present two approaches named ``Top-Down'' and ``Bottom-Up''. The underlying idea revolves around the structured nature of file metadata, which forms a tree-like hierarchy from the highest level (import libraries) to the lowest level (executable section blocks), encompassing lists of functionalities. Exploiting this structure, we aim to identify the resilient and invariant path of similarities, starting from the top and progressing towards the bottom. In the subsequent subsections, we delve into both approaches, providing comprehensive descriptions of the algorithms employed. Furthermore, we present the results obtained from our investigations.





\subsection{Top-Down Approach}
To detect and evaluate potential resilient fingerprints, the Top-down approach first clusters resilient fingerprints based on 100\% similarities in their import libraries and functionalities. Next, we conduct a detailed analysis of the sections in these resilient fingerprints, with a particular focus on identifying any malicious or camouflage sections. Based on this analysis, we create a comprehensive profile for each resilient fingerprint, including detailed statistics at both the file and section levels. By analyzing these statistics, we can determine the likelihood that a given file belongs to a resilient fingerprint.

Algorithm~\ref{algo1:top-down} outlines the top-down process. It takes two separate parameters, $\texttt\it{{feeds}}$ and $\texttt\it{{sections}}$, and groups the feeds by their import list similarity $\texttt\it{{imphash}}$. The groups are then sorted in descending order based on their count, allowing us to focus on the significant clusters first (steps 1-2). Next, we iterate through each cluster in $\texttt\it{{SortimpLists}}$, using each entry's $\texttt\it{{imphash}}$ to locate all relevant files from $\texttt\it{{feeds}}$. We then calculate file-level statistics, such as entropy and hashes (steps 4-6). We also dive into $\texttt\it{{sections}}$ to retrieve all the sections that belong to the resilient fingerprint, group those sections using $\texttt\it{{secHash}}$, and sort them in descending order based on count to prioritize significant sections (steps 7-9). We then iterate through each cluster of sections within $\texttt\it{{sortsecKeys}}$, using each entry's $\texttt\it{{secKey}}$ to locate all relevant sections before collecting all the statistics about that common section (steps 12-13). For example, if we have ten sections with similar keys, we collect the minimum and maximum entropy, and minimum and maximum chi-squared distance (Chi2). We decide to focus on the top 10 sections within each resilient fingerprint as a threshold. Finally, we combine the collected $\texttt\it{{feedsInfo}}$ and $\texttt\it{{section[]}}$ to create $\texttt\it{{resilientFprintInfo}}$, which forms the basis of our decision-making process.

\begin{algorithm}[!ht]
\scriptsize
\caption{Top-Down Approach}\label{algo1:top-down}
\begin{algorithmic}[1]
\Statex \textbf{Input: } $\texttt{feeds},\texttt{sections}$ \Comment{$feeds: f_1,f_2..,f_n. sections: s_1,s_2..,s_n$} 
\Statex \textbf{Output: } $\texttt{resilientFprintInfo}$ \Comment{set of resilient fingerprints, each has related meta-info}
\State $\texttt{impLists}$ = group \texttt{feeds} by \texttt{imphash}  \Comment{imphash is unique identifier for import list}
\State $\texttt{SortimpLists}$ = sort $\texttt{impLists}$ by count in descending order  
\State $\texttt{resilientFprintInfo}$ = []  
\For{$\texttt{imphash}$ in $\texttt{SortimpLists}$}
\State $\texttt{resilientFingerprints{[]}}$ = locate $\texttt{feeds}$  by $\texttt{imphash}$
\State $\texttt{feedsInfo}$ = calculate $\texttt{resilientFingerprints{[]}}$  stats
\State $\texttt{RFprintSections}$ = locate relevant $\texttt{sections}$  of $\texttt{resilientFingerprints{[]}}$
\State $\texttt{secKeys}$ = group $\texttt{RFprintSections}$ by $\texttt{secHash}$  \Comment{secHash is unique section identifier}
\State $\texttt{sortsecKeys}$ = sort $\texttt{secKeys}$ descendingly by count  
\State $\texttt{secInfo}$ = [] 
\For{$\texttt{secKey}$ in $\texttt{sortsecKeys}$}
\State $\texttt{section{[]}}$ = locate $\texttt{RFprintSections}$  by $\texttt{secKey}$
\State $\texttt{sparams}$ = calculate $\texttt{section{[]}}$  stats
\State append $\texttt{sparams}$ to $\texttt{secInfo}$
\EndFor
\State append $\texttt{feedsInfo}$,$\texttt{section{[]}}$,$\texttt{secInfo{[]}}$ to  $\texttt{resilientFprintInfo}$
\EndFor
\end{algorithmic}
\end{algorithm}

\subsection{Bottom-Up Approach}
This approach is going the other way around to start from the bottom up by clustering all files that use common sections, including camouflage or malicious. This approach is motivated by our analysis, where \textit{we find that common camouflage sections and malicious sections could be found in another resilient fingerprint that has different import list libraries when we follow the Top-Bottom approach}. We attribute that to  Malware authors slightly manipulating the import list libraries to avoid being detected or using the same executable sections for different intent. Therefore, starting from the bottom (sections) up, we can capture resilient fingerprints robust to slight import variations. Similar to the previous approach, a profile is built to determine the maliciousness of the fingerprints.

Algorithm~\ref{algo2:bottom-up} outlines the bottom-up process. It takes two similar separate parameters, $\texttt\it{{feeds}}$ and $\texttt\it{{sections}}$ of the entire group (1M feeds and their sections). We begin by diving into $\texttt\it{{sections}}$ to retrieve all common sections, group those sections using $\texttt\it{{secHash}}$, and sort them in descending order based on count to prioritize significant sections (steps 1-2). We then iterate through each cluster of sections within $\texttt\it{{sortsecKeys}}$, using each entry's $\texttt\it{{secKey}}$ to locate all relevant sections (steps 4-5). Next, we use collected $\texttt\it{{RFprintSections[]}}$ to go up to the file level and retrieve all relevant file-level feeds (steps 6). We observe here that we might encounter redundant file-level feeds as multiple sections might belong to the same file and cause double dipping counting. Hence, we introduce step 7 as a filtering condition to eliminate this redundancy using $\texttt\it{{feedID}}$ before collecting all the statistics about that common section (steps 8-9).

\begin{algorithm}[!h]
\scriptsize
\caption{Bottom-Up Approach}\label{algo2:bottom-up}
\begin{algorithmic}[1]
\Statex \textbf{Input:} $\texttt{feeds},\texttt{sections}$ \Comment{$feeds: f_1,f_2..,f_n. sections: s_1,s_2..,s_n$}
\Statex \textbf{Output:} $\texttt{resilientFprintInfo}$\Comment{set of resilient fingerprints, each  has related meta-info}
\State $\texttt{secKeys}$ = group $\texttt{sections}$ by $\texttt{secHash}$  \Comment{secHash is unique identifier}
\State $\texttt{sortsecKeys}$ = sort $\texttt{secKeys}$ descendingly by count
\State $\texttt{resilientFprintInfo}$ = []  
\For{$\texttt{secKey}$ in $\texttt{sortsecKeys}$}
\State $\texttt{RFprintSections{[]}}$ = locate $\texttt{sections}$  by $\texttt{secKey}$
\State $\texttt{resilientFprint{[]}}$ = locate $\texttt{feeds}$  by $RFprintSections{[]}$
\State $\texttt{resilientFprint{[]}}$ = drop duplicates $resilientFprint{[]}$  by $\texttt{feedsID}$
\State $\texttt{sparams}$ = calculate $\texttt{RFprintSections{[]}}  \text{stats}$
\State $\texttt{fparams}$ = calculate $\texttt{resilientFprint{[]}} \text{stats}$
\State append $\texttt{sparams}$,$\texttt{fparams}$ to  $\texttt{resilientFprintInfo}$
\EndFor
\end{algorithmic}
\end{algorithm}

\subsection{Results}
\subheading{Overview and settings.} As detailed in Algorithm ~\ref{algo1:top-down} and \ref{algo2:bottom-up}, we obtain all the required statistical heuristics from both approaches. In the results section, we explore various combinations across the four groups of our datasets to examine the potential efficacy. Those combinations include Import List (IL), Redundant Sections (RS), Camouflage Sections (CS), and Malicious Sections (MS) as they have been defined in Invariant Parts (Section~\ref{sec:invariation}). We use the vendor's labels as ground truth to evaluate the maliciousness of the identified fingerprints. As stated early, we select a conservative threshold ($t>=4$) as an indicator to consider a given binary as malicious. We also refer to the number of files within a resilient fingerprint as redundancy. Lastly, the accuracy column denotes the effectiveness of our approach in identifying files and assessing their probability of being associated with a resilient fingerprint, rather than  conducting binary classification between benign and malware.    

We identify three types of resilient fingerprints. The first is fully malicious, meaning four or more vendors flag all the identified files within that resilient fingerprint as malicious. The second type is resilient fingerprints where all the files are flagged by less than four vendors, and we mark them as False Positive (FP). The third and most interesting resilient fingerprints type is partial malicious, meaning some of the files are highly flagged by the vendors, but others are not while they share common import lists and camouflage or malicious section blocks. We combine those resilient fingerprints toward True Positive (TP) accuracy. The results below will provide more details and a specific focus on the  FP and TP. Note that all thresholding has been derived from group 1 only. Therefore, we use groups 2, 3, and 4 to examine the generalisability of our findings.\\ 

\subheading{Top-Down Results.}
We analyze the Top-Down approach using five combinations across four groups. The results in Table~\ref{tb:top-downDetection} are as follows. (1) IL and RS combination: Identify approximately 560 fingerprints with a high FP rate (around 6.8\%) and a TP rate of about 93.1\%. Consistent results are observed across all four groups, with slightly higher FP (9\%) and lower TP (90\%). 
(2) IL and CS combination: Reduce the number of identified fingerprints by more than half, resulting in improved FP rates (between 1.3\% and 2.8\%) and TP rates (between 97.1\% and 98.6\%). Identify around 25\% of 1 million files within each group.
(3) IL and MS combination: Increase the total number of identified resilient fingerprints to two-thirds. Reasonable FP rates (between 2.2\% and 3.9\%) and TP rates (between 96\% and 97.7\%) are achieved, identifying around 323K-434K of all 1 million files within each group.
(4) IL, CS, and MS combination: Among all combinations, it has one of the lowest numbers of identified resilient fingerprints (between 195-347). However, it provides decent FP rates (between 1.3\% and 2.8\%) and TP rates (between 97.1\% and 98.6\%), identifying between 213K-227K of all files within each group. This combination is considered too strict.
(5) Based on these observations, the fifth combination of IL and either CS or MS proves to be the most effective. It identifies the highest number of resilient fingerprints (between 345 and 1025) with FP rates (between 2.2\% and 3.9\%) and TP rates (between 96\% and 97.7\%). It identifies slightly more files (between 324K-339K), representing approximately 33\% of all 1 million files within each group. Thus, the presence of IL, combined with either CS or MS, is deemed sufficient to connect a file to a resilient fingerprint. 

\begin{table}
\centering
\caption{Top-Down Approach resilient Fingerprinting.}
\label{tb:top-downDetection}
\vspace{-0.3cm}
\resizebox{3.4in}{!}{
\begin{tabular}{|c|c|c|c|c|c|c|c|c|} 
\hline
\multirow{2}{*}{Top-Down}                & \multirow{2}{*}{\begin{tabular}[c]{@{}c@{}}\\ Fingerprint\end{tabular}} & \multicolumn{3}{c|}{\begin{tabular}[c]{@{}c@{}}\\ (F\textless{}4 )\end{tabular}} & \begin{tabular}[c]{@{}c@{}}\\ ( F\textless{}4,F\textgreater{}=4 )\end{tabular} & \begin{tabular}[c]{@{}c@{}}\\ (F\textgreater{}=4)\end{tabular} & \multicolumn{2}{c|}{\begin{tabular}[c]{@{}c@{}}\\ (TP)\end{tabular}}  \\ 
\cline{3-9}
                                         &                                                                         & Num & Acc.(\%) & Redundancy                                                           & \multicolumn{2}{c|}{Num}                                                                                                                        & Acc.(\%) & Redundancy                                                      \\ 
\hline
\multicolumn{9}{|c|}{Group 1}                                                                                                                                                                                                                                                                                                                                                                                                   \\ 
\hline
ILRS                                     & 560                                                                     & 51  & 6.8      & 40,034                                                          & 158                                                                            & 351                                                            & 93.1     & 544,613                                                    \\ 
\hline
ILCS                                     & 222                                                                     & 11  & 1.3      & 3,070                                                           & 67                                                                             & 144                                                            & 98.6     & 224,803                                                    \\ 
\hline
ILMS                                     & 394                                                                     & 32  & 2.2      & 7,572                                                           & 103                                                                            & 261                                                            & 97.7     & 335,062                                                    \\ 
\hline
ILCSMS                                   & 219                                                                     & 11  & 1.3      & 3,070                                                           & 66                                                                             & 142                                                            & 98.6     & 223,513                                                    \\ 
\hline
\rowcolor[rgb]{0.867,1,0.867} ILCS(or)MS & 345                                                                     & 22  & 2.2     & 7,572                                                           & 95                                                                             & 231                                                            & 97.7     & 339,278                                                    \\ 
\hline
\multicolumn{9}{|c|}{Group 2}                                                                                                                                                                                                                                                                                                                                                                                                   \\ 
\hline
ILRS                                     & 478                                                                     & 56  & 9.0      & 53,682                                                          & 148                                                                            & 274                                                            & 90.9     & 540,439                                                    \\ 
\hline
ILCS                                     & 197                                                                     & 13  & 2.0      & 4,407                                                           & 67                                                                             & 117                                                            & 97.9     & 213,656                                                    \\ 
\hline
ILMS                                     & 344                                                                     & 39  & 2.9      & 9,993                                                           & 97                                                                             & 208                                                            & 97.0     & 323,728                                                    \\ 
\hline
ILCSMS                                   & 195                                                                     & 13  & 2.0      & 4,407                                                           & 66                                                                             & 116                                                            & 97.9     & 213,021                                                    \\ 
\hline
\rowcolor[rgb]{0.867,1,0.867} ILCS(or)MS & 346                                                                     & 39  & 2.9      & 9,993                                                           & 98                                                                             & 209                                                            & 97.0     & 324,363                                                    \\ 
\hline
\multicolumn{9}{|c|}{Group 3}                                                                                                                                                                                                                                                                                                                                                                                                   \\ 
\hline
ILRS                                     & 494                                                                     & 62  & 9.3      & 56,581                                                          & 144                                                                            & 288                                                            & 90.6     & 548,525                                                    \\ 
\hline
ILCS                                     & 196                                                                     & 17  & 2.8      & 6,499                                                           & 61                                                                             & 118                                                            & 97.1     & 218,264                                                    \\ 
\hline
ILMS                                     & 355                                                                     & 43  & 3.4      & 11,753                                                          & 93                                                                             & 219                                                            & 96.5     & 329,789                                                    \\ 
\hline
ILCSMS                                   & 195                                                                     & 17  & 2.8      & 6,499                                                           & 61                                                                             & 117                                                            & 97.1     & 217,782                                                    \\ 
\hline
\rowcolor[rgb]{0.867,1,0.867} ILCS(or)MS & 356                                                                     & 43  & 3.4      & 11,753                                                          & 93                                                                             & 220                                                            & 96.6   & 330,271                                                    \\ 
\hline
\multicolumn{9}{|c|}{Group 4 (Collected 9 months after)}                                                                                                                                                                                                                                                                                                                                                                        \\ 
\hline
ILRS                                     & 1419                                                                    & 88  & 4.3      & 29,309                                                          & 164                                                                            & 1167                                                           & 95.6     & 652,168                                                    \\ 
\hline
ILCS                                     & 351                                                                     & 15  & 2.7      & 6,403                                                           & 15                                                                             & 258                                                            & 97.2     & 229,597                                                    \\ 
\hline
ILMS                                     & 1021                                                                    & 60  & 3.9      & 17,678                                                          & 127                                                                            & 834                                                            & 96.0     & 434,068                                                    \\ 
\hline
ILCSMS                                   & 347                                                                     & 14  & 2.6      & 6,236                                                           & 78                                                                             & 255                                                            & 97.3     & 227,134                                                    \\ 
\hline
\rowcolor[rgb]{0.867,1,0.867} ILCS(or)MS & 1025                                                                    & 61  & 3.9      & 17,845                                                          & 127                                                                            & 837                                                            & 96.0     & 436,531                                                    \\
\hline
\end{tabular}}
\end{table}

\begin{table}
\centering
\caption{Bottom-Up Approach resilient Fingerprinting.}
\label{tb:bottom-upDetection}
\vspace{-0.3cm}
\resizebox{3.4in}{!}{
\begin{tabular}{|c|c|c|c|c|c|c|c|c|} 
\hline
\multirow{2}{*}{Bottom-Up}             & \multirow{2}{*}{\begin{tabular}[c]{@{}c@{}}\\ Fingerprint\end{tabular}} & \multicolumn{3}{c|}{\begin{tabular}[c]{@{}c@{}}\\ (F\textless{}4 )\end{tabular}} & \begin{tabular}[c]{@{}c@{}}\\ ( F\textless{}4,F\textgreater{}=4 )\end{tabular} & \begin{tabular}[c]{@{}c@{}}\\ (F\textgreater{}=4)\end{tabular} & \multicolumn{2}{c|}{\begin{tabular}[c]{@{}c@{}}\\ (TP)\end{tabular}}  \\ 
\cline{3-9}
                                       &                                                                         & Num  & Acc.  (\%) & Redundancy                                                        & \multicolumn{2}{c|}{Num}                                                                                                                        & Acc. (\%) & Redundancy                                                     \\ 
\hline
\multicolumn{9}{|c|}{Group 1}                                                                                                                                                                                                                                                                                                                                                                                                 \\ 
\hline
RS                                     & 603                                                                     & 141  & 5.1        & 34,640                                                       & 187                                                                            & 278                                                            & 94.9      & 656,219                                                   \\ 
\hline
CS                                     & 26                                                                      & 4    & 0.1        & 7                                                            & 21                                                                             & 1                                                              & 99.9      & 457,365                                                   \\ 
\hline
MC                                     & 273                                                                     & 39   & 4.1        & 4,038                                                        & 67                                                                             & 167                                                            & 95.9      & 105,719                                                   \\ 
\hline
\rowcolor[rgb]{0.867,1,0.867} CS(or)MC & 299                                                                     & 43   & 0.8        & 4,448                                                        & 88                                                                             & 168                                                            & 99.2      & 563,084                                                   \\ 
\hline
\multicolumn{9}{|c|}{Group 2}                                                                                                                                                                                                                                                                                                                                                                                                 \\ 
\hline
RS                                     & 511~                                                                    & ~151 & 6.1        & 43,071~                                                      & 164~                                                                           & 196~                                                           & 93.9      & ~665,236                                                  \\ 
\hline
CS                                     & 27~                                                                     & 1~   & ~0.1       & ~2                                                           & 26~                                                                            & 0~                                                             & 100~      & ~467,425                                                  \\ 
\hline
MS                                     & ~232                                                                    & 47~  & 5.1        & ~6,987                                                       & ~54                                                                            & ~131                                                           & ~94.9     & 132,010~                                                  \\ 
\hline
\rowcolor[rgb]{0.867,1,0.867} CS(or)MS & ~259                                                                    & ~48  & 1.2        & 6,987~                                                       & ~80                                                                            & 131~                                                           & ~98.8     & ~599,435                                                  \\ 
\hline
\multicolumn{9}{|c|}{Group 3}                                                                                                                                                                                                                                                                                                                                                                                                 \\ 
\hline
RS                                     & 534                                                                     & 139  & 6.3        & 42,984                                                       & 163                                                                            & 226                                                            & 93.7      & 648,942                                                   \\ 
\hline
CS                                     & 32                                                                      & 3    & 0.1        & 217                                                          & 27                                                                             & 1                                                              & 99.9      & 462,836                                                   \\ 
\hline
MS                                     & 248                                                                     & 40   & 4.5        & 5,511                                                        & 54                                                                             & 153                                                            & 95.5      & 118,403                                                   \\ 
\hline
\rowcolor[rgb]{0.867,1,0.867} CS(or)MS & 280                                                                     & 43   & 0.9        & 5,728                                                        & 81                                                                             & 154                                                            & 99.1      & 581,239                                                   \\ 
\hline
\multicolumn{9}{|c|}{Group 4 (Collected 9 months after)}                                                                                                                                                                                                                                                                                                                                                                      \\ 
\hline
RS                                     & 1,177                                                                   & 146  & 3.3        & 25,803                                                       & 196                                                                            & 838                                                            & 96.7      & 758,426                                                   \\ 
\hline
CS                                     & 52                                                                      & 4    & 0.1        & 73                                                           & 39                                                                             & 11                                                             & 99.9      & 451,284                                                   \\ 
\hline
MS                                     & 607                                                                     & 43   & 4.2        & 7,575                                                        & 60                                                                             & 505                                                            & 95.8      & 173,785                                                   \\ 
\hline
\rowcolor[rgb]{0.867,1,0.867} CS(or)MS & 659                                                                     & 47   & 1.3        & 7,648                                                        & 99                                                                             & 516                                                            & 98.7      & 625,069                                                   \\
\hline
\end{tabular}}
\end{table}

\subheading{Bottom-Up Results.} Table~\ref{tb:bottom-upDetection} presents the results of our bottom-up approach using four combinations across the four groups. (1) Connection based on redundant sections: This method identifies the most resilient fingerprints (511-1177) with a TP rate of 93\%-96\%, but has a relatively high FP rate (3\%-6\%). It captures about 650K files (65\%) from the four groups. (2) Connection based on camouflage sections: While identifying fewer fingerprints (26-52), this approach captures around 450K files (45\%) with a very low FP rate (0.016\%) and an impressive TP rate of 99.9\%. (3) Connection based on malicious sections: This method identifies 232-273 fingerprints, with a TP rate of 95\%, but has a higher FP rate (4\%-5\%). It captures the fewest files (100K-175K or 10\%-17.5\%) out of 1 million. (4) Combined approach: Combining CS and MC yields the best results, identifying 280-659 resilient fingerprints with a low FP rate (1\%) and a TP rate of >98.7\%. It captures 563K-625K files (50\%-60\%) across all 1 million files in each group.

\subheading{Comparison with Traditional Fingerprinting.} 
We compare our best results from the Top-Down and Bottom-Up approaches, highlighted in green, against two traditional fingerprinting methods: SHA256 and TLSH. For the Top-Down approach, we select the best-performing combination of IL with either CS or MS, while for the Bottom-Up approach, we choose the best CS or MC combinations. Among cryptographic hashes (e.g., MD5, SHA1, SHA256), we select SHA256 for its widespread use. All three cryptographic hashes perform similarly across the groups. For fuzzy hashes, we select TLSH due to its speed and scalability, which have led to its adoption by major platforms like VirusTotal, Malware Bazaar, MISP, and STIX~\cite{oliver2013tlsh}.

Table~\ref{tb:comparision} shows that SHA256 had the lowest accuracy (16\%-18.4\%) in detecting file similarities across the four groups, calculated based on hash redundancies. TLSH performed slightly better, with accuracy ranging from 17\% to 20.7\%. In contrast, our Top-Down and Bottom-Up approaches outperform both traditional methods. The Top-Down approach, using Import Lists, camouflage sections, and malicious sections, achieves 30\%-41.9\% accuracy, while the Bottom-Up approach achieves the best performance with 53\%-58\%. Additionally, our approaches maintain low FP rates, with the Top-Down approach at 2.2\%-3.9\% and the Bottom-Up approach at around 1\%, outperforming traditional fingerprinting in both accuracy and false positives.

\begin{table}
\centering
\caption{Comparison with Tradition Fingerprinting.}
\label{tb:comparision}
\vspace{-0.3cm}
\resizebox{3.2in}{!}{
\begin{tabular}{|c|c|c|c|c|c|c|c|c|} 
\hline
\multirow{2}{*}{\bf Technique} & \multicolumn{2}{c|}{\begin{tabular}[c]{@{}c@{}}{\bf Group 1} \\1,061,151\end{tabular}} & \multicolumn{2}{c|}{\begin{tabular}[c]{@{}c@{}}{\bf Group 2} \\1,061,151\end{tabular}} & \multicolumn{2}{c|}{\begin{tabular}[c]{@{}c@{}}{\bf Group 3} \\1,061,151\end{tabular}} & \multicolumn{2}{c|}{\begin{tabular}[c]{@{}c@{}}{\bf Group 4}\\1,061,151\end{tabular}}  \\ 
\cline{2-9}
                           & \# of files & Acc.~                                                              & \# of files & Acc.                                                               & \# of files & Acc.                                                               & \# of files & Acc.                                                               \\ 
\hline
SHA256                     & 172,102     & 16.2                                                               & 184,592     & 17.4                                                               & 169,891     & 16.0                                                               & 195,655     & 18.4                                                               \\ 
\hline
TLSH                       & 195,028     & 18.4                                                               & 210,027     & 19.8                                                               & 189,493     & 17.9                                                               & 219,304     & 20.7                                                               \\ 
\hline
Top-Down                   & 339,278     & 31.9                                                               & 324,363     & 30.5                                                               & 330,271     & 31.1                                                               & 436,365     & 41.1                                                               \\ 
\hline
Bottom-Up                  & 563,084     & 53.1                                                               & 599,435     & 56.4                                                               & 581,239     & 54.7                                                               & 625,069     & 58.9                                                               \\
\hline
\end{tabular}}
\end{table}

\observ{\small We propose two novel mechanisms, the Top-Down and Bottom-Up approaches, to enhance malware fingerprinting. These methods consistently perform well across four datasets, successfully identifying 30\% to 58\% of reported suspicious files. This represents a significant improvement, up to tripling the identification rate of traditional fingerprinting mechanisms}
\section{Discussion and Limitations}

\subheading{Top-Down vs. Bottom-Up Approaches.}
After conducting systematic evaluations and answering Q1 and Q2, we identified several invariant parts within files that could improve traditional malware fingerprinting techniques, including cryptographic hash-based and fuzzy hash-based methods. To enhance the detection of similar malware files with common tactics such as import lists, camouflage sections, and malicious sections, we developed two novel techniques: Top-Down and Bottom-Up.
While both methods proved effective, we noted some trade-offs. Bottom-Up outperformed Top-Down in terms of detection accuracy, but it tended to capture partially malicious resilient fingerprints with more files, whereas Top-Down leaned toward capturing fully malicious resilient fingerprints. We argue that Top-Down may be the better option when a security analyst aims to capture the resilient fingerprints' intent. This approach begins with common Import libraries with a list of functionalities and then delves into the camouflage and malicious sections, providing a clear view of the intent. In contrast, Bottom-Up may be less effective at identifying the resilient fingerprints' intent but more suitable for capturing a larger number of malware files that do not necessarily share the same import libraries but use camouflage and malicious executable sections tactics.\\


\subheading{False Positive Fingerprints.} Our work intentionally avoided using predictive models like neural networks and instead relied on statistical heuristics to identify connected malware files in the early stages of triaging. While our methods do have a false positive rate, it is very low. For instance, the Top-Down approach has a 2-3.9\% false positive rate, whereas the Bottom-Up approach has an even lower false positive rate of 0.7-1.2\%. It's worth noting that not all the files within false positive deep fingerprints have zero malicious vendor labels. Many of them have been labeled malicious by 1-3 vendors, but our threshold for counting a file as malicious is a minimum of 4 vendor labels, pushing these files into the false positive cluster. Additionally, there is a possibility that some of these files are highly malicious, but vendors did not detect them when we collected the dataset. Unfortunately, we were unable to confirm this observation.

\subheading{Evaluation of Good Executables.} Utilizing benign executables sourced externally to obtain VirusTotal reports presented a significant challenge, primarily stemming from IP and legal concerns rather than technical obstacles. Given the focus of our study on Windows PE files, initiating the submission of benign Windows executables to VirusTotal as suspicious for metadata generation would raise ethical issues. However, our collection of samples from VirusTotal already includes a substantial percentage of benign files. As indicated in Table~\ref{tb:benignLabels}, approximately 25\% to 35\% of samples in each group exhibit benign characteristics with \textit{zero}  malicious flags. We believe this proportion is deemed satisfactory for our research objectives.

\subheading{Packing and Generating Reports Overhead.} We emphasize, in Figure 1, that the VirusTotal platform includes packing type details (e.g., UPX) within the metadata parameters. To this end,  VirusTotal routinely unpacks these samples before generating metadata. This deliberate omission from our discussion is rooted in the scope of our work. In terms of generation overhead, our study leveraged VirusTotal for metadata generation. Every minute, they supplied metadata for approximately 1,500 received files. However, we selectively utilized detailed metadata, focusing on quick static analysis of sections and import lists as full entities. Typically, these processes demand minimal time, often measured in seconds.

\subheading{Limitations and Future Directions.} Our work has two main limitations. Firstly, although our approach incorporates more flexible fingerprinting mechanisms to analyze file-invariant parts, we still rely on the hash of these parts as a conservative measure. However, an adaptive attacker could potentially slightly alter these parts, necessitating the use of similarity techniques like cosine similarity with an appropriate threshold instead of content hashing. Secondly, our Top-Down approach heavily depends on import libraries' list of functionalities and content hash for clustering. However, a small percentage of reports collected from VirusTotal lack this content, possibly due to obfuscated import libraries. To address this, we could utilize API deobfuscation models such as~\cite{cheng2021obfuscation} to rebuild the Import libraries list before calculating their content hash, mitigating this limitation.

\section{Related Work}
The concept of ``malware fingerprinting'' encompasses two distinct approaches. Firstly, it involves utilizing a defense mechanism that calculates a hash value from suspicious files and shares it publicly to identify similar files at an early stage, which is the focus of our current work. Alternatively, it may entail observing the behavior of malware during execution to detect its presence. The second approach involves an evasion mechanism whereby malware attempts to detect signs of analysis environments or debuggers, as detailed in a study by Afianian et al.~\cite{afianian2019malware}.\\

\subheading{Traditional Fingerprinting.} 
Stream hashing is a key technique in malware and threat intelligence, serving as the initial line of defense against cyber threats. It is widely utilized and shared among security teams as checksums or unique identifiers for triage purposes, with platforms like VirusTotal, MalwareBazaar, MISP, and STIX~\cite{tlsh} being popular for this purpose. Security analysts use these identifiers (hashes) to label and share malware across teams.
Two main mechanisms generate these identifiers: cryptographic based hashing and fuzzy-based hashing. Cryptographic hashing, such as MD5, SHA-1, and SHA-256, uses cryptographic diffusion to hide the relationship between the original entity and the hash, but even minor changes in the entity result in a completely different hash~\cite{goldmaxMS}.

To address this limitation, fuzzy hashing offers a more flexible approach by tolerating minor changes between files, making it useful for capturing some malware evasion techniques. Examples include Nilsimsa, TLSH, SSDEEP, and SDHASH, with TLSH being the most widely adopted due to its speed and scalability. However, our findings indicate that while fuzzy hashes outperform cryptographic hashes, their effectiveness remains low (<20\%) at the file level, underscoring the need for more adaptable fingerprinting strategies to identify diverse file groups.

\subheading{Dynamic Fingerprinting.} Another work stream focuses on fingerprinting malware based on its execution behavior. Studies by Willems et al.~\cite{willems2007toward} and Rossow et al.~\cite{rossow2011sandnet} found that executing malware for a short period accurately identifies its fingerprint and malicious intent. Kilgallon et al.~\cite{kilgallon2017improving} proposed a systematic mechanism for predicting the time required for sufficient malicious behavior. Küchler et al.
~\cite{kuchler2021does} conducted a study on malware behavior in sandboxes and developed a machine learning-based detection method. They find that multiple sandbox runs may be necessary for reliable behavior fingerprinting. However, this approach may not be suitable for the initial triaging of thousands of suspicious files.

\subheading{Systematic Studies.} Several recent large-scale studies have examined different aspects of malware evasion. Barr-Smith et al.~\cite{barr2021survivalism} focused on the Living off the land technique and found detection gaps among the top 10 AntiVirus products. Alrawi et al.~\cite{alrawi2021circle} compared the lifecycle of IoT malware with traditional desktop-targeting malware and highlighted differences in infection vectors and command and control communication. Cozzi et al.~\cite{cozzi2018understanding} studied Linux-targeting malware and observed similar tactics to Windows malware. Duan~\cite{duan2018things} conducted a study on Android packers using whole-system emulation. Wong et al.~\cite{yong2021inside} proposed a malware analysis taxonomy and recommended guidelines for developers. 

However, to the best of our knowledge, no large-scale systematic study has explored the limitations of traditional malware fingerprinting for detection and the need for more adaptable mechanisms to identify mutated files with common intent, which motivated the current work.

\section{Conclusion}
This paper addresses the gap in understanding how malware authors bypass traditional file-level fingerprinting and proposes novel approaches to enhance it. We conducted a large-scale empirical analysis of Windows PE files from VirusTotal, identifying invariant and evasive variant parts. Our two proposed approaches, Top-Down and Bottom-Up, cluster files using resilient fingerprinting strategies based on similarities in their Import Libraries and sections. The results show a potential improvement of over 50\% in detecting similar malware files with variations compared to traditional methods. These findings underscore the need to update malware detection fingerprinting methods to counter evolving evasion techniques.


\balance
\bibliographystyle{unsrt}
\bibliography{ref}

\begin{thebibliography}{10}

\bibitem{anderson2018ember}
Hyrum~S Anderson and Phil Roth.
\newblock Ember: an open dataset for training static pe malware machine
  learning models.
\newblock {\em arXiv preprint arXiv:1804.04637}, 2018.

\bibitem{downing2021deepreflect}
Evan Downing, Yisroel Mirsky, Kyuhong Park, and Wenke Lee.
\newblock Deepreflect: Discovering malicious functionality through binary
  reconstruction.
\newblock In {\em USENIX Security Symposium}, pages 3469--3486, 2021.

\bibitem{mariconti2016mamadroid}
Enrico Mariconti, Lucky Onwuzurike, Panagiotis Andriotis, Emiliano
  De~Cristofaro, Gordon Ross, and Gianluca Stringhini.
\newblock Mamadroid: Detecting android malware by building markov chains of
  behavioral models.
\newblock {\em Proceedings of 24th Network and Distributed System Security
  Symposium (NDSS)}, 2017.

\bibitem{jindal2019neurlux}
Chani Jindal, Christopher Salls, Hojjat Aghakhani, Keith Long, Christopher
  Kruegel, and Giovanni Vigna.
\newblock Neurlux: dynamic malware analysis without feature engineering.
\newblock In {\em Proceedings of the 35th Annual Computer Security Applications
  Conference}, pages 444--455, 2019.

\bibitem{wang2020you}
Qi~Wang, Wajih~Ul Hassan, Ding Li, Kangkook Jee, Xiao Yu, Kexuan Zou, Junghwan
  Rhee, Zhengzhang Chen, Wei Cheng, Carl~A Gunter, et~al.
\newblock You are what you do: Hunting stealthy malware via data provenance
  analysis.
\newblock In {\em NDSS}, 2020.

\bibitem{IDAPro}
IDAPro.
\newblock The interactive disassembler pro.
\newblock \url{https://hex-rays.com/ida-pro/}, 2023.

\bibitem{sikorski2012practical}
Michael Sikorski and Andrew Honig.
\newblock {\em Practical malware analysis: the hands-on guide to dissecting
  malicious software}.
\newblock no starch press, 2012.

\bibitem{oliver2021designing}
Jonathan Oliver and Josiah Hagen.
\newblock Designing the elements of a fuzzy hashing scheme.
\newblock In {\em 2021 IEEE 19th International Conference on Embedded and
  Ubiquitous Computing (EUC)}, pages 1--6. IEEE, 2021.

\bibitem{virustotal}
Virustotal.
\newblock \url{https://www.virustotal.com/}, 2023.

\bibitem{tlsh}
Tlsh - a locality sensitive hash.
\newblock \url{https://tlsh.org/}, 2021.

\bibitem{goldmaxMS}
Andrea~Lelli Ramin~Nafisi.
\newblock Goldmax, goldfinder, and sibot: Analyzing nobelium’s layered
  persistence.
\newblock
  \url{https://www.microsoft.com/en-us/security/blog/2021/03/04/goldmax-goldfinder-sibot-analyzing-nobelium-malware/},
  2021.

\bibitem{barr2021survivalism}
Frederick Barr-Smith, Xabier Ugarte-Pedrero, Mariano Graziano, Riccardo
  Spolaor, and Ivan Martinovic.
\newblock Survivalism: Systematic analysis of windows malware
  living-off-the-land.
\newblock In {\em 2021 IEEE Symposium on Security and Privacy (SP)}, pages
  1557--1574. IEEE, 2021.

\bibitem{alrawi2021circle}
Omar Alrawi, Charles Lever, Kevin Valakuzhy, Ryan Court, Kevin~Z Snow, Fabian
  Monrose, and Manos Antonakakis.
\newblock The circle of life: A large-scale study of the iot malware lifecycle.
\newblock In {\em USENIX Security Symposium}, pages 3505--3522, 2021.

\bibitem{yao2023hiding}
Mingxuan Yao, Jonathan Fuller, Ranjita~Pai Kasturi, Saumya Agarwal, Amit~Kumar
  Sikder, and Brendan Saltaformaggio.
\newblock Hiding in plain sight: an empirical study of web application abuse in
  malware.
\newblock In {\em 32nd USENIX Security Symposium (USENIX Security 23)}, pages
  6115--6132, 2023.

\bibitem{cozzi2018understanding}
Emanuele Cozzi, Mariano Graziano, Yanick Fratantonio, and Davide Balzarotti.
\newblock Understanding linux malware.
\newblock In {\em 2018 IEEE symposium on security and privacy (SP)}, pages
  161--175. IEEE, 2018.

\bibitem{yong2021inside}
Miuyin Yong~Wong, Matthew Landen, Manos Antonakakis, Douglas~M Blough, Elissa~M
  Redmiles, and Mustaque Ahamad.
\newblock An inside look into the practice of malware analysis.
\newblock In {\em Proceedings of the 2021 ACM SIGSAC Conference on Computer and
  Communications Security}, pages 3053--3069, 2021.

\bibitem{sabir2022reliability}
Bushra Sabir, M~Ali Babar, Raj Gaire, and Alsharif Abuadbba.
\newblock Reliability and robustness analysis of machine learning based
  phishing url detectors.
\newblock {\em IEEE Transactions on Dependable and Secure Computing}, 2022.

\bibitem{shmalko2022profiler}
Mariya Shmalko, Alsharif Abuadbba, Raj Gaire, Tingmin Wu, Hye-Young Paik, and
  Surya Nepal.
\newblock Profiler: Distributed model to detect phishing.
\newblock In {\em 2022 IEEE 42nd International Conference on Distributed
  Computing Systems (ICDCS)}, pages 1336--1337. IEEE, 2022.

\bibitem{mousavi2024investigation}
Zahra Mousavi, Chadni Islam, Kristen Moore, Alsharif Abuadbba, and M~Ali Babar.
\newblock An investigation into misuse of java security apis by large language
  models.
\newblock In {\em Proceedings of the 19th ACM Asia Conference on Computer and
  Communications Security}, pages 1299--1315, 2024.

\bibitem{abuadbba2022towards}
Alsharif Abuadbba, Shuo Wang, Mahathir Almashor, Muhammed~Ejaz Ahmed, Raj
  Gaire, Seyit Camtepe, and Surya Nepal.
\newblock Towards web phishing detection limitations and mitigation.
\newblock {\em arXiv preprint arXiv:2204.00985}, 2022.

\bibitem{evans2022raider}
Keelan Evans, Alsharif Abuadbba, Tingmin Wu, Kristen Moore, Mohiuddin Ahmed,
  Ganna Pogrebna, Surya Nepal, and Mike Johnstone.
\newblock Raider: Reinforcement-aided spear phishing detector.
\newblock In {\em International Conference on Network and System Security},
  pages 23--50. Springer, 2022.

\bibitem{vt_scanapi}
Virustotal: Upload and scan a file api.
\newblock \url{https://developers.virustotal.com/reference/}, 2023.

\bibitem{invernizzi2014nazca}
Luca Invernizzi, Stanislav Miskovic, Ruben Torres, Christopher Kruegel,
  Sabyasachi Saha, Giovanni Vigna, Sung-Ju Lee, and Marco Mellia.
\newblock Nazca: detecting malware distribution in large-scale networks.
\newblock In {\em NDSS}, volume~14, pages 23--26, 2014.

\bibitem{kolodenker2017paybreak}
Eugene Kolodenker, William Koch, Gianluca Stringhini, and Manuel Egele.
\newblock Paybreak: Defense against cryptographic ransomware.
\newblock In {\em Proceedings of the 2017 ACM on Asia Conference on Computer
  and Communications Security}, pages 599--611, 2017.

\bibitem{kotzias2016measuring}
Platon Kotzias, Leyla Bilge, and Juan Caballero.
\newblock Measuring $\{$PUP$\}$ prevalence and $\{$PUP$\}$ distribution through
  $\{$Pay-Per-Install$\}$ services.
\newblock In {\em 25th USENIX Security Symposium (USENIX Security 16)}, pages
  739--756, 2016.

\bibitem{sharif2018predicting}
Mahmood Sharif, Jumpei Urakawa, Nicolas Christin, Ayumu Kubota, and Akira
  Yamada.
\newblock Predicting impending exposure to malicious content from user
  behavior.
\newblock In {\em Proceedings of the 2018 ACM SIGSAC conference on computer and
  communications security}, pages 1487--1501, 2018.

\bibitem{sarabi2018characterizing}
Armin Sarabi and Mingyan Liu.
\newblock Characterizing the internet host population using deep learning: A
  universal and lightweight numerical embedding.
\newblock In {\em Proceedings of the Internet Measurement Conference 2018},
  pages 133--146, 2018.

\bibitem{schwartz2018using}
Edward~J Schwartz, Cory~F Cohen, Michael Duggan, Jeffrey Gennari, Jeffrey~S
  Havrilla, and Charles Hines.
\newblock Using logic programming to recover c++ classes and methods from
  compiled executables.
\newblock In {\em Proceedings of the 2018 ACM SIGSAC Conference on Computer and
  Communications Security}, pages 426--441, 2018.

\bibitem{thomas2016investigating}
Kurt Thomas, Juan A~Elices Crespo, Ryan Rasti, Jean-Michel Picod, Cait
  Phillips, Marc-Andr{\'e} Decoste, Chris Sharp, Fabio Tirelo, Ali Tofigh,
  Marc-Antoine Courteau, et~al.
\newblock Investigating commercial $\{$Pay-Per-Install$\}$ and the distribution
  of unwanted software.
\newblock In {\em 25th USENIX Security Symposium (USENIX Security 16)}, pages
  721--739, 2016.

\bibitem{wong2018tackling}
Michelle~Y Wong and David Lie.
\newblock Tackling runtime-based obfuscation in android with $\{$TIRO$\}$.
\newblock In {\em 27th USENIX Security Symposium (USENIX Security 18)}, pages
  1247--1262, 2018.

\bibitem{wang2014whowas}
Liang Wang, Antonio Nappa, Juan Caballero, Thomas Ristenpart, and Aditya
  Akella.
\newblock Whowas: A platform for measuring web deployments on iaas clouds.
\newblock In {\em Proceedings of the 2014 Conference on Internet Measurement
  Conference}, pages 101--114, 2014.

\bibitem{tian2018needle}
Ke~Tian, Steve~TK Jan, Hang Hu, Danfeng Yao, and Gang Wang.
\newblock Needle in a haystack: Tracking down elite phishing domains in the
  wild.
\newblock In {\em Proceedings of the Internet Measurement Conference 2018},
  pages 429--442, 2018.

\bibitem{wwwchain19}
Muhammad Ikram, Rahat Masood, Gareth Tyson, Mohamed~Ali Kaafar, Noha Loizon,
  and Roya Ensafi.
\newblock The chain of implicit trust: An analysis of the web third-party
  resources loading.
\newblock In {\em The World Wide Web Conference}, pages 2851--2857, 2019.

\bibitem{zhao2019decade}
Benjamin Zi~Hao Zhao, Muhammad Ikram, Hassan~Jameel Asghar, Mohamed~Ali Kaafar,
  Abdelberi Chaabane, and Kanchana Thilakarathna.
\newblock A decade of mal-activity reporting: A retrospective analysis of
  internet malicious activity blacklists.
\newblock In {\em Proceedings of the 2019 ACM Asia Conference on Computer and
  Communications Security}, pages 193--205, 2019.

\bibitem{cai2016inferring}
Zhenquan Cai and Roland~HC Yap.
\newblock Inferring the detection logic and evaluating the effectiveness of
  android anti-virus apps.
\newblock In {\em Proceedings of the Sixth ACM Conference on Data and
  Application Security and Privacy}, pages 172--182, 2016.

\bibitem{korczynski2017capturing}
David Korczynski and Heng Yin.
\newblock Capturing malware propagations with code injections and code-reuse
  attacks.
\newblock In {\em Proceedings of the 2017 ACM SIGSAC Conference on Computer and
  Communications Security}, pages 1691--1708, 2017.

\bibitem{zhu2020measuring}
Shuofei Zhu, Jianjun Shi, Limin Yang, Boqin Qin, Ziyi Zhang, Linhai Song, and
  Gang Wang.
\newblock Measuring and modeling the label dynamics of online
  $\{$Anti-Malware$\}$ engines.
\newblock In {\em 29th USENIX Security Symposium (USENIX Security 20)}, pages
  2361--2378, 2020.

\bibitem{song2016learning}
Linhai Song, Heqing Huang, Wu~Zhou, Wenfei Wu, and Yiying Zhang.
\newblock Learning from big malwares.
\newblock In {\em Proceedings of the 7th ACM SIGOPS Asia-Pacific Workshop on
  Systems}, pages 1--8, 2016.

\bibitem{oliver2013tlsh}
Jonathan Oliver, Chun Cheng, and Yanggui Chen.
\newblock Tlsh--a locality sensitive hash.
\newblock In {\em 2013 Fourth Cybercrime and Trustworthy Computing Workshop},
  pages 7--13. IEEE, 2013.

\bibitem{cheng2021obfuscation}
Binlin Cheng, Ming Jiang, Erika Leal, Haotian Zhang, Jianming Fu, Guojun Peng,
  and Jean-Yves Marion.
\newblock Obfuscation-resilient executable payload extraction from packed
  malware.
\newblock In {\em 30th Usenix Security Sympoisum}, 2021.

\bibitem{afianian2019malware}
Amir Afianian, Salman Niksefat, Babak Sadeghiyan, and David Baptiste.
\newblock Malware dynamic analysis evasion techniques: A survey.
\newblock {\em ACM Computing Surveys (CSUR)}, 52(6):1--28, 2019.

\bibitem{willems2007toward}
Carsten Willems, Thorsten Holz, and Felix Freiling.
\newblock Toward automated dynamic malware analysis using cwsandbox.
\newblock {\em IEEE Security \& Privacy}, 5(2):32--39, 2007.

\bibitem{rossow2011sandnet}
Christian Rossow, Christian~J Dietrich, Herbert Bos, Lorenzo Cavallaro, Maarten
  Van~Steen, Felix~C Freiling, and Norbert Pohlmann.
\newblock Sandnet: Network traffic analysis of malicious software.
\newblock In {\em Proceedings of the First Workshop on Building Analysis
  Datasets and Gathering Experience Returns for Security}, pages 78--88, 2011.

\bibitem{kilgallon2017improving}
Sean Kilgallon, Leonardo De~La~Rosa, and John Cavazos.
\newblock Improving the effectiveness and efficiency of dynamic malware
  analysis with machine learning.
\newblock In {\em 2017 Resilience Week (RWS)}, pages 30--36. IEEE, 2017.

\bibitem{kuchler2021does}
Alexander K{\"u}chler, Alessandro Mantovani, Yufei Han, Leyla Bilge, and Davide
  Balzarotti.
\newblock Does every second count? time-based evolution of malware behavior in
  sandboxes.
\newblock In {\em NDSS}, 2021.

\bibitem{duan2018things}
Yue Duan, Mu~Zhang, Abhishek~Vasisht Bhaskar, Heng Yin, Xiaorui Pan, Tongxin
  Li, Xueqiang Wang, and XiaoFeng Wang.
\newblock Things you may not know about android (un) packers: A systematic
  study based on whole-system emulation.
\newblock In {\em NDSS}, 2018.

\end{thebibliography}

\clearpage

\appendix

\section{Ground Truth of top 18 identified Resilient fingerprints}\label{appendix:18fingerprints}

\begin{figure}[h]
     \centering
     \begin{subfigure}[b]{0.7\textwidth}
         \centering
         \includegraphics[width=\textwidth]{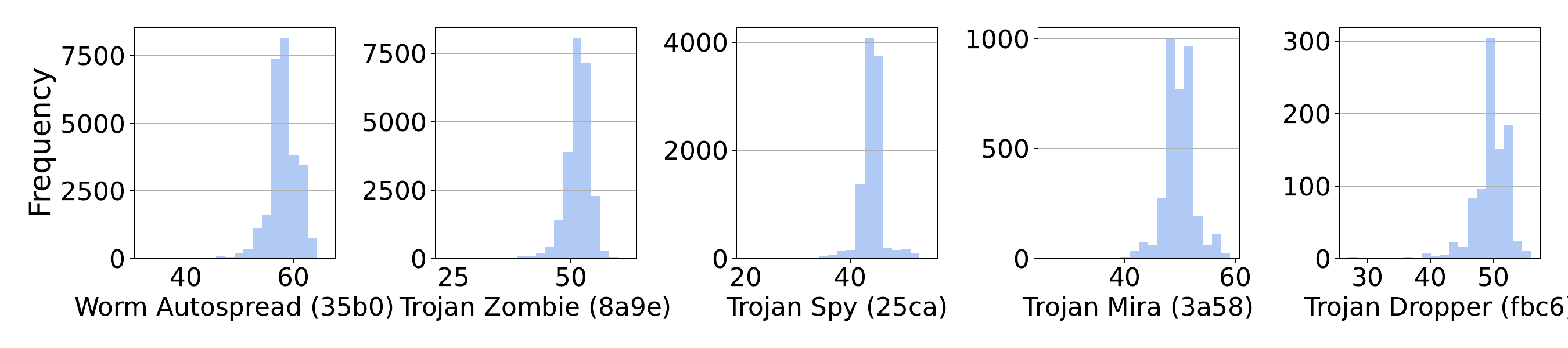}
         \caption{Group 1}
         \label{fig:y equals x}
     \end{subfigure}
     \hfill
     \begin{subfigure}[b]{0.7\textwidth}
         \centering
         \includegraphics[width=\textwidth]{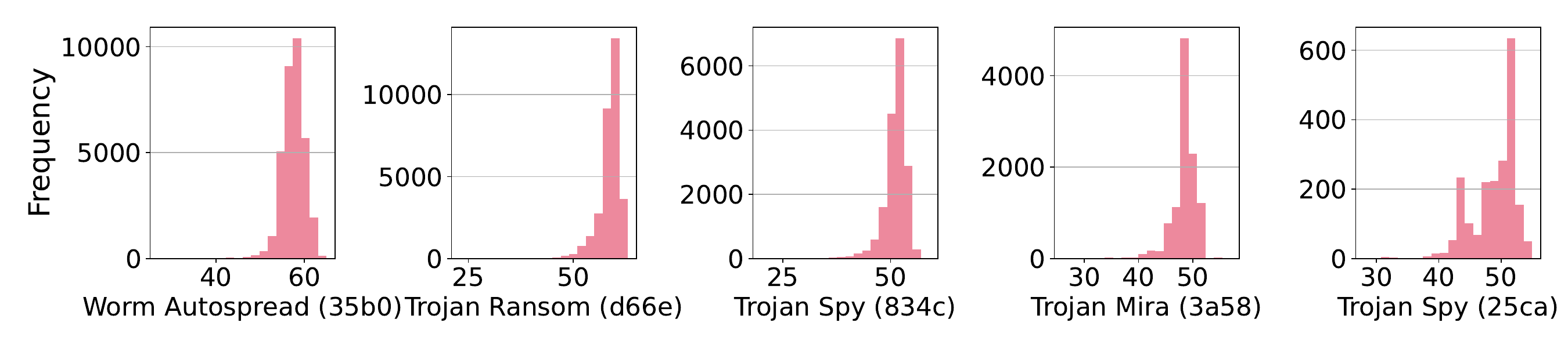}
         \caption{Group 2}
         \label{fig:three sin x}
     \end{subfigure}
     \hfill
     \begin{subfigure}[b]{0.7\textwidth}
         \centering
         \includegraphics[width=\textwidth]{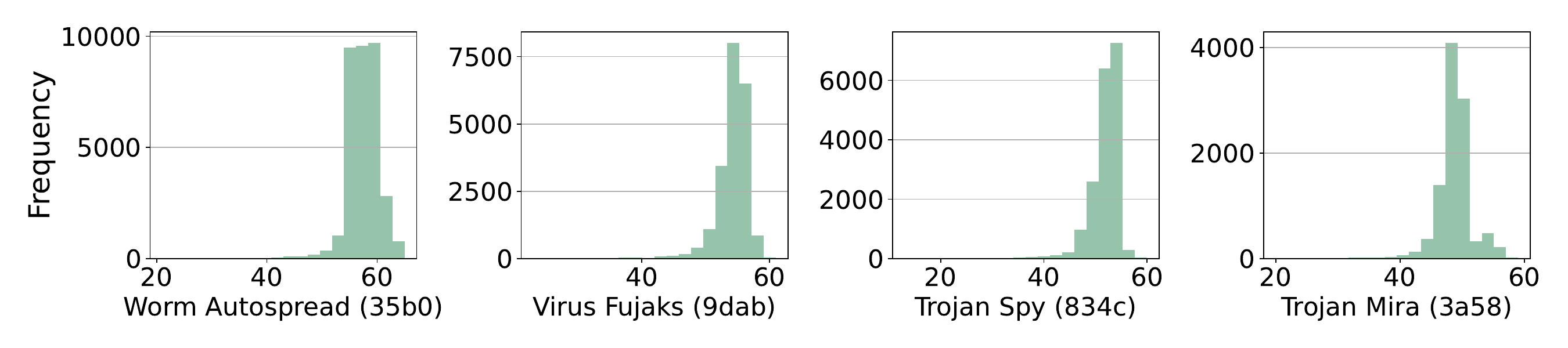}
         \caption{Group 3}
         \label{fig:five over x}
     \end{subfigure}
     \begin{subfigure}[b]{0.7\textwidth}
         \centering
         \includegraphics[width=\textwidth]{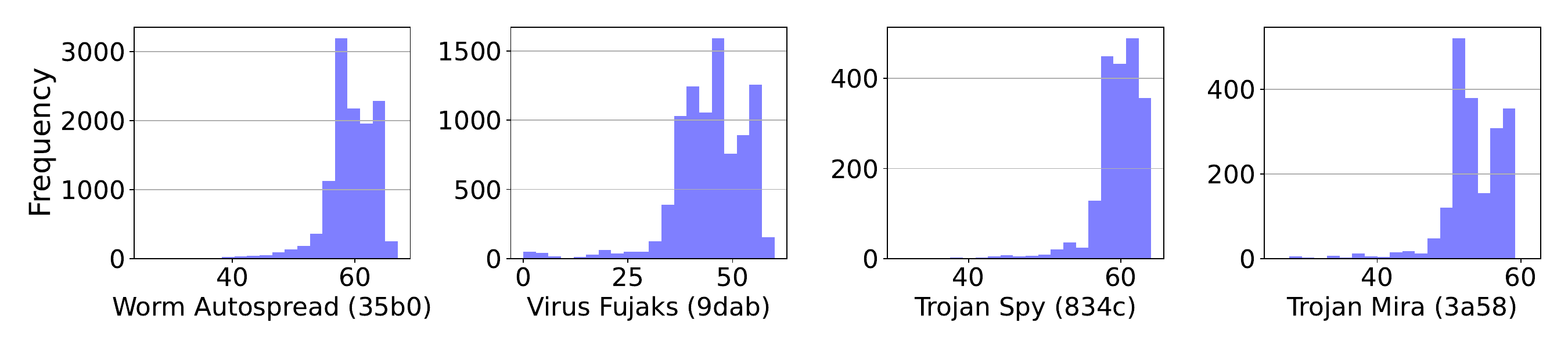}
         \caption{Group 4}
         \label{fig:three sin x}
     \end{subfigure}
        \caption{ Ground Truth for the extracted campaigns from
Groups 1, 2, 3 and 4.}
        \label{fig:groundTruthG}
\end{figure}

Figure~\ref{fig:groundTruthG} shows on the x-axis that all files within each resilient fingerprint are highly malicious, as they have been labelled as malicious by  $>$20 vendors.

\section{VirusTotal vendors malicious labels and file types distribution} \label{appendix:vt_filedistribution}

Table~\ref{tb:benignLabels} shows the malicious label distributions of VirusTotal vendors, which also shows between 25\%-35\% are benign with zero malicious flags. Also, Table~\ref{tb:filetype} shows the file-type distributions in our study.

\begin{table}[h]
\centering
\caption{VirusTotal vendors malicious labels. When 0,1,2 or 3 vendors flag as malicious, we consider as within the benign boundaries.}
\resizebox{2.2in}{!}{
\label{tb:benignLabels}
\begin{tabular}{c|c|c|c|c|c} 
\hline
\multirow{2}{*}{Dataset} & \multicolumn{4}{c|}{Files Malicous Flags}                                                                                                                                                                                                & \multirow{2}{*}{Total}                                     \\ 
\cline{2-5}
                         & 0                                                          & 1                                                       & 2                                                       & 3                                                       &                                                            \\ 
\hline
Group1                   & \begin{tabular}[c]{@{}c@{}}349,635\\(32.9\%)\end{tabular}  & \begin{tabular}[c]{@{}c@{}}50,737\\(4.8\%)\end{tabular} & \begin{tabular}[c]{@{}c@{}}21,265\\(2.0\%)\end{tabular} & \begin{tabular}[c]{@{}c@{}}12,679\\(1.2\%)\end{tabular} & \begin{tabular}[c]{@{}c@{}}434,316\\(40.9\%)\end{tabular}  \\ 
\hline
Group2                   & \begin{tabular}[c]{@{}c@{}}344,259\\(32.4\%)\end{tabular}  & \begin{tabular}[c]{@{}c@{}}59,838\\(5.6\%)\end{tabular} & \begin{tabular}[c]{@{}c@{}}29,448\\(2.8\%)\end{tabular} & \begin{tabular}[c]{@{}c@{}}16,871\\(1.6\%)\end{tabular} & \begin{tabular}[c]{@{}c@{}}450,416\\(42.4\%)\end{tabular}  \\ 
\hline
Group3                   & \begin{tabular}[c]{@{}c@{}}376,198\\(35.5\%)\end{tabular}  & \begin{tabular}[c]{@{}c@{}}52,207\\(4.9\%)\end{tabular} & \begin{tabular}[c]{@{}c@{}}21,169\\(2.0\%)\end{tabular} & \begin{tabular}[c]{@{}c@{}}11,550\\(1.1\%)\end{tabular} & \begin{tabular}[c]{@{}c@{}}461,124\\(43.5\%)\end{tabular}  \\ 
\hline
Group4                   & \begin{tabular}[c]{@{}c@{}}272,897 \\(25.7\%)\end{tabular} & \begin{tabular}[c]{@{}c@{}}60,702\\(5.7\%)\end{tabular} & \begin{tabular}[c]{@{}c@{}}23,549\\(2.2\%)\end{tabular} & \begin{tabular}[c]{@{}c@{}}11,597\\(1.1\%)\end{tabular} & \begin{tabular}[c]{@{}c@{}}368,745\\(34.5\%)\end{tabular}  \\
\hline
\end{tabular}}
\end{table}

\begin{table}[h]
\centering
\caption{File Types distribution of four groups.}
\centering
\label{tb:filetype}
\resizebox{2.8in}{!}{
\begin{tabular}{c|c|c|c|c|c|c|c|c} 
\hline
          & \multicolumn{2}{c|}{Group 1} & \multicolumn{2}{c|}{Group 2} & \multicolumn{2}{c|}{Group 3} & \multicolumn{2}{c}{Group 4}  \\ 
\hline
File Type & Count   & \%                 & Count   & \%                 & Count   & \%                 & Count   & \%                  \\ 
\hline
Win32 EXE & 666,583 & 62.8               & 643,955 & 60.6               & 612,417 & 57.7               & 801,322 & 75.5                \\ 
\hline
Win32 DLL & 207,323 & 19.5               & 221,605 & 20.8               & 202,667 & 19.1               & 130,598 & 12.3                \\ 
\hline
Win64 EXE & 104,251 & 9.8                & 105,580 & 9.9                & 95,553  & 9.0                & 67,292  & 6.3                 \\ 
\hline
Win64 DLL & 82,899  & 7.8                & 80,012  & 7.5                & 77,285  & 7.2                & 61,844  & 5.8                 \\
\hline
\end{tabular}}
\end{table}

\section{Unreliable Invariant non-Executable Parts Example}\label{appendix:unreliableParts}
\begin{figure}[h]
    \centering
    \includegraphics[width=0.6\linewidth]{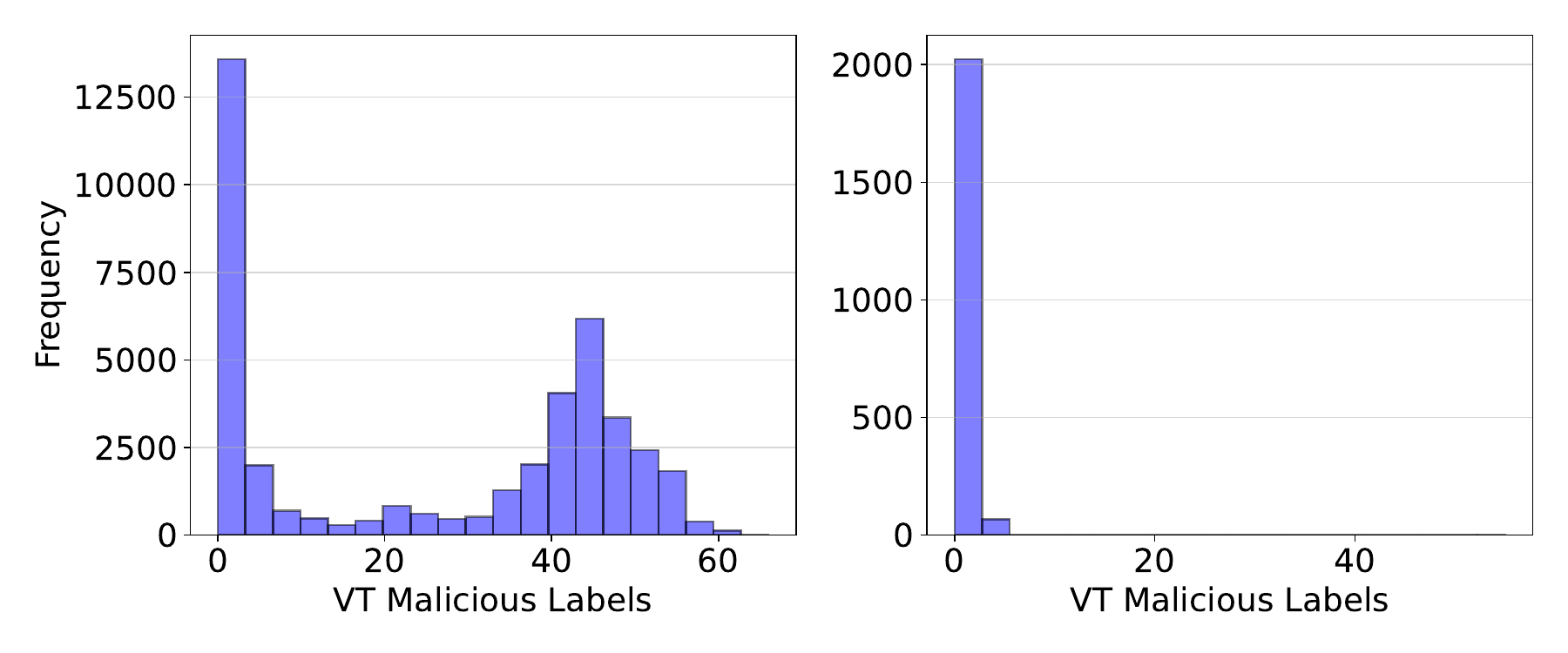}
    \caption{Ground truth for two extracted fingerprints based on the import list high level and resources as low level.}
    \label{fig:resourcesCampaign}
\end{figure}

Figure~\ref{fig:resourcesCampaign} illustrates the distribution of maliciousness across two noteworthy identified resilient fingerprints based on the import list and common resources, consisting of  41,578 and 2,094 files, respectively.  The first resilient fingerprint demonstrates a suspicious trend, as some of its files are highly malicious while others are considered benign. In contrast, the second resilient fingerprint is primarily labeled as benign. These findings suggest that using resources as a key similarity tactic may not be a reliable mechanism for identifying resilient fingerprints.

\section{Variation of Section Names}\label{appendix:sectionNames}

Figure~\ref{fig:secion_names} illustrates that the section names, which have the same content with 10,295 files, show significant section name randomness, indicating they were likely automatically generated as a tactic to conceal their true purpose.

\begin{figure}[h]
    \centering
    \includegraphics[width=0.3\linewidth]{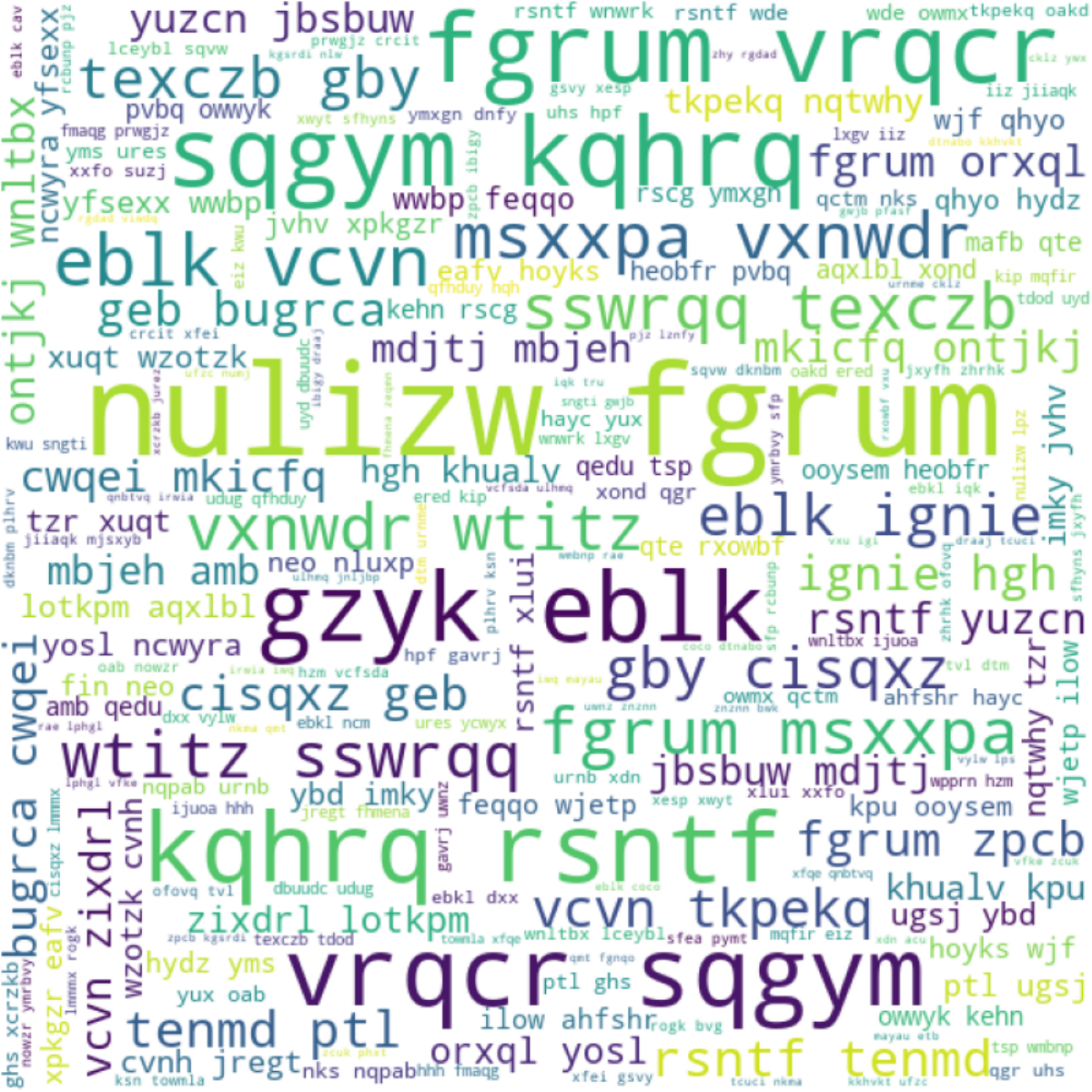}
    \caption{Same section content across 10,295 files with variant section names.}
    \label{fig:secion_names}
    \vspace{-4mm}
\end{figure}

\end{document}